\documentclass[fleqn,usenatbib]{mnras}

\usepackage{newtxtext,newtxmath}
\usepackage{graphicx}

\usepackage[T1]{fontenc}

\DeclareRobustCommand{\VAN}[3]{#2}
\let\VANthebibliography\thebibliography
\def\thebibliography{\DeclareRobustCommand{\VAN}[3]{##3}\VANthebibliography}

\usepackage{graphicx}	
\usepackage{amsmath}	
\usepackage{amsfonts}
\usepackage{orcidlink}

\newcommand{\alper}{$\alpha$ Per }

\definecolor{color1}{rgb}{0.20,0.48,0.72} 
\definecolor{figblue}{RGB}{77, 76, 187} 
\definecolor{figred}{RGB}{178, 51, 42} 
\definecolor{teal}{RGB}{42, 160, 140} 
\definecolor{amber}{RGB}{180, 145, 60} 

\newcommand{\addcite}[1]{\textcolor{orange}{cite}}

\newcommand{\Msun}{\mbox{M$_{\odot}$}}

\title[NCS VI - Rotation in PLATO's LOPS2 field]{NGTS clusters survey - VI: Stellar rotation in seven young open clusters within the PLATO LOPS2 field}

\author[A. Hughes et al.]{
Alexander Hughes$^{1}$\thanks{E-mail: alex.hughes@qmul.ac.uk}\orcidlink{0009-0001-3561-5035},
Edward Gillen$^{1}$\orcidlink{0000-0003-2851-3070},
Matthew Battley$^{1}$\orcidlink{0000-0002-1357-9774},
Deepak Chahal$^{1}$\orcidlink{0000-0002-3612-3622},
Beatrice Caccherano$^{1}$\orcidlink{0009-0001-3865-0119},
\newauthor
David Anderson$^{2}$,
Ioannis Apergis$^{3,4}$\orcidlink{0009-0004-7473-4573},
Daniel Bayliss$^{3}$\orcidlink{0000-0001-6023-1335},
Matthew Burleigh$^{5}$\orcidlink{0000-0003-0684-7803},
Jorge Fern\'andez Fern\'andez$^{3,4}$\orcidlink{0000-0002-1416-2188},
\newauthor
Michael Goad$^{5}$,
George Harvey$^{5}$,
James S. Jenkins$^{7,8}$\orcidlink{0000-0003-2733-8725},
Alicia Kendall$^{5}$,
James McCormac$^{3,4}$\orcidlink{0000-0003-1631-4170},
\newauthor
Jose Moyano$^{2}$,
Gavin Ramsay$^{6}$\orcidlink{0000-0001-8722-9710},
Suman Saha$^{7,8}$\orcidlink{0000-0001-8018-0264},
Jose Vines$^{2}$\orcidlink{0000-0002-2135-9018},
Richard G. West$^{3,4}$\orcidlink{0000-0001-6604-5533},
Peter J. Wheatley$^{3,4}$\orcidlink{0000-0003-1452-2240}
\\
$^{1}$Astronomy Unit, Queen Mary University of London, Mile End Road, London E1 4NS, UK\\
$^{2}$Instituto de Astronom\'ia, Universidad Cat\'olica del Norte, Angamos 0610, 1270709, Antofagasta, Chile\\
$^{3}$Department of Physics, University of Warwick, Gibbet Hill Road, Coventry CV4 7AL, UK\\
$^{4}$Centre for Exoplanets and Habitability, University of Warwick, Gibbet Hill Road, Coventry CV4 7AL, UK\\
$^{5}$School of Physics \& Astronomy, University of Leicester, Leicester LE1 7RH, UK\\
$^{6}$Armagh Observatory \& Planetarium, College Hill, Armagh, BT61 9DG, UK\\
$^{7}$Instituto de Estudios Astrof\'isicos, Facultad de Ingenier\'ia y Ciencias, Universidad Diego Portales, Av. Ej\'ercito Libertador 441, Santiago, Chile \\
$^{8}$Centro de Excelencia en Astrof\'isica y Tecnolog\'ias Afines (CATA), Camino El Observatorio 1515, Las Condes, Santiago, Chile \\
}

\date{Accepted XXX. Received YYY; in original form ZZZ}

\pubyear{\the\year{}}

\begin{document}
\label{firstpage}
\pagerange{\pageref{firstpage}--\pageref{lastpage}}
\maketitle

\begin{abstract}
We present NGTS measurements of rotation period distributions for FGKM stars in seven young open clusters spanning $\sim$\,40--700\,Myr within PLATO's first long-stare LOPS2 field. We measure 1063 rotation periods, of which 479 are newly analysed as part of cluster specific rotation studies, whilst 63 are unique periods not reported in the recent TESS All-Sky Rotation Survey. Of the 1063 rotation periods, 285 are identified as likely binary or higher order multiple systems using colour-magnitude diagrams and \textit{Gaia} astrometry. These are the first comprehensive rotation period distributions for Trumpler 10, NGC 2451\,B and Alessi 3, while extending existing distributions for NGC 2451\,A, NGC 2516, Collinder 135 and IC 2391, to create a fuller picture on the rotation state of young stars in PLATO's LOPS2 field. We find that main-sequence solar-mass stars in the $\sim$\,40\,Myr old NGC 2451\,B cluster, form a slow sequence that can be distinguished from their counterparts at $\sim$\,70--80\,Myr, thereby significantly reducing the age at which young stellar groups can be relatively aged via their rotation sequences. We also observe stalled spin down from the age of NGC 2451\,A to at least that of NGC 2516 ($\sim$\,70\,--\,150\,Myr) at masses $\gtrsim1$ \Msun{}, supporting previous predictions that angular momentum redistribution and removal should result in a wave of stalled spin down that propagates as a function of both mass and age. Finally, we provide a new age estimate for Alessi 3 of 687$\,\pm$\,106 Myr using differential gyrochronology age dating. 
\end{abstract}




\section{Introduction}
Angular momentum, together with mass and composition, determines much of a star's evolutionary pathway. 
At the start of the pre-main sequence (PMS), stars are observed to have rotation periods spanning $\sim$1--10 days \citep{2020AJ....159..273R,2023MNRAS.523..169S}. Magnetic star-disk interactions transfer angular momentum, leading to disk locking that can prevent a star from spinning up as it contracts \citep{1991ApJ...370L..39K,2005ApJ...634.1214L,2017A&A...599A..23V,2018AJ....155..196R}. Once the inner disk begins to disperse, the star continues to spin up as it contracts, until it arrives at the zero-age main sequence (ZAMS) where contraction is halted by the onset of fusion in the core. The distribution of disk lifetimes is therefore widely accepted as the origin of the spread of rotation periods observed at the ZAMS \citep{2013A&A...556A..36G, 2015A&A...577A..98G}, with stars that undergo shorter disk-locking emerging as faster rotators in rotation-mass space.\par

PMS lifetimes are mass-dependent with higher mass stars reaching the ZAMS earlier, i.e. $\sim$\,15, 40 and 160 Myr for a 1.5, 1 and 0.5 $M_{\sun}$ star respectively \citep{MIST2}. For stars $\gtrsim 0.35$\,\Msun{} \citep{1997A&A...327.1039C}, PMS contraction leads to the formation of a radiative core that decouples from the convective envelope, allowing differential rotation to develop between the two layers. The timescale over which angular momentum is redistributed between them to restore solid body rotation is known as the core-envelope coupling timescale ($\tau_\mathrm{c}$; \citealt{1991ApJ...376..204M, 1998A&A...333..629A, 2013A&A...556A..36G}). This timescale plays a major role in the angular momentum and subsequent rotation period evolution around the ZAMS, with lower mass stars having significantly longer coupling timescales than solar type stars and rapid rotators having coupling timescales 3-6 times shorter than slow rotators \citep{2015A&A...577A..98G}. Short coupling timescales allow for efficient transportation of angular momentum between the core and envelope, further spinning up rapid rotators. On the main sequence, magnetized stellar winds cause the star to spin down over time \citep[e.g.][]{1995ApJ...441..865C, 1995ApJ...441..876C,2012ApJ...746...43R, 2021ApJ...912...65G}, the magnitude of which is proportional to $\Omega^3$ where $\Omega$ is the angular velocity of the stellar envelope \citep{1988ApJ...333..236K}. Thus, once on the ZAMS, fast rotators experience a larger braking force relative to a slow rotator of the same mass and converge onto a well defined sequence (typically referred to as the slow, or sometimes converged, sequence) which increases in period monotonically as a function of age.\par

This mass-dependent change in rotation period as a function of time means that the converged slow sequence can be exploited to age-date stars, a process known as gyrochronology \citep{1972ApJ...171..565S, 2003ApJ...586..464B, 2007ApJ...669.1167B}. With the uptick in available data over the last few decades from wide-field photometric surveys \citep[eg; Kepler, K2, TESS, NGTS;][]{kepler,k2,tess,2018MNRAS.475.4476W} recent studies have turned to open clusters with well-defined ages to calibrate gyrochronal models. Photometric observations of stars allow inference of the rotation period by monitoring the modulation in received flux, typically resulting from star spots on the stellar surface moving through the line of sight inline with the stellar surface rotational velocity. By leveraging these observations with data on stellar mass (or some appropriate proxy), it is possible to constrain the period distribution at a given mass and age to calibrate gyrochronal models. \par

Recent studies have looked at this in great detail for young clusters (<1 Gyr) to try and calibrate gyrochronal models during key evolutionary periods \citep[eg;][]{2010MNRAS.408..475H,2016AJ....152..113R,2019AJ....158...77C,2020A&A...641A..51F, 2020MNRAS.492.1008G, 2021ApJ...921..167R, 2022A&A...657L...3M, 2023MNRAS.523..169S,  2024ApJ...962...16D, chronoflow}. \cite{2023ApJ...947L...3B} found that age uncertainties typically reduce with age, with uncertainty on solar mass stars improving monotonically between $0.2$\,--\,$2$\,Gyr. Recent work by \cite{2023AJ....166...14B} has provided a new lower anchor for benchmark clusters used to calibrate gyrochronal models, however at $\sim 80$\,Myr, this is still much later than the ZAMS for solar-type stars. Whilst \cite{2024ApJ...962...16D} recently increased the number of measured periods for solar mass stars by an order of magnitude, the overall population size of young stars with measured rotation periods is still small. Because gyrochronal models are often empirically driven, studying young clusters is imperative if we are to calibrate models at important ages such as the ZAMS, which in turn helps to better understand the level of scatter exhibited by the population due to intrinsic (e.g. rotation, magnetic fields) and extrinsic (e.g. disc lifetime, environment) factors. One upcoming mission that aims to help with this is the Planetary Transits and Oscillations of Stars \citep[PLATO;][]{2014ExA....38..249R}. Planned for launch in early 2027, PLATO's Long-duration Observation Phase (LOP) plans to observe a region of the southern sky centered at $l= 255.9375\degr$ \& $b= -24.62432\degr$ known as LOPS2 \citep{2025A&A...694A.313N}. With a $\sim 49\degr \times 49\degr$ field of view (FoV), LOPS2 covers roughly 5\% of the sky and will be observed for a minimum of two years \citep{2024arXiv240605447R}. One of PLATO's main science objectives is to use the light curves attained through long stare observations to conduct a comprehensive gyrochronal study, calibrating gyrochronal models through a range of important evolutionary ages and furthering our understanding of early stellar, and subsequently planetary evolution. \par

In this paper, we use observations with the Next-Generation Transit Survey (NGTS) to measure stellar rotation periods for stars in seven clusters within the LOPS2 field: Alessi 3, Trumpler 10,  Collinder 135, NGC2516, NGC 2451\,A-B and IC2391. For three of these clusters, the results of this work represent the first known dedicated rotation study: Alessi 3, Trumpler 10 \& NGC 2451\,B. We also draw from available literature for rotation studies of clusters in this work to provide a comprehensive rotation survey prior to the launch of PLATO. Five of the clusters in this study also exist at ages around the ZAMS for solar-type stars. Section \ref{sec cl data} of the paper outlines the observations and clusters analysed in this work, before providing the methodology used to attain our results, along with a literature comparison\footnote{Section \ref{sec measuring prot} outlines the methodology undertaken in this work only and does not cover the methodology of literature studies referenced within this work during rotation sequence comparison.} in section \ref{sec measuring prot}. We discuss the results of this study in Section \ref{sec disc}, before concluding with section \ref{sec conc}. 

\section{Observations and Cluster Characterisation}
\label{sec cl data}

\subsection{Cluster membership}
\label{ssec ngts clusters}
To determine cluster membership, we conducted a literature review to find all studies which had compiled membership catalogues of open clusters. Using these studies we created a master list for each cluster which fell fully or partially in the PLATO field by including any star that was included in at least one of these literature catalogues. 

This literature review only included membership studies conducted after the release of Gaia Data Release 2 \citep[DR2; ][]{babusiaux18_ext} because of the significantly increased precision and reliability of Gaia astrometric data compared to earlier instruments \citep{Gaia2016,babusiaux18_ext}. In total, this literature review revealed 19 individual catalogues which ranged from studies of individual clusters \citep{Damiani2019,Dickson-Van2020,Gagne2018_VC,Gagne2020,Galli2020,Kounkel2018,Luhman2018,Luhman2020b,Villa-Velez2018} to wider clustering studies considering many clusters at once \citep{Cantat-Gaudin2018,Cantat-Gaudin2019a,Cantat-Gaudin2019b,Cantat-Gaudin2020,Gagne2018,2018A&A...618A..93C,Kounkel-Covey2019,Kuhn2021,Meingast2021,Zari2018}.\footnote{Note this also included information from the tess-infos repository; https://github.com/MNGuenther/tess\_infos} Because each study approaches their membership criteria independently, they may be sensitive to slightly different populations of stars. To maximise survey completeness while retaining only the most reliable members of each cluster, we include a star as a member if its membership probability is $\geq70$\% in at least one catalogue. Note however that almost all targets are identified as a member by multiple studies, reducing the chance of spurious membership and systematic biases whilst retaining as many high probability members as possible.

One additional consideration is that of filamentary structures, colloquially referred to as tidal tails or stellar streams. The study conducted by \citet{Kounkel-Covey2019} made use of a clustering method called Hierarchical Density-Based Spatial Clustering of Applications with Noise, \citep[HDBSCAN;][]{McInnes2017}, with their analysis identifying extended filamentary structures at younger ages. Some of these identified filaments contain previously known clusters as part of the strings substructure. Later work by \citet{Meingast2021} questioned the physical relevance of these structures, stating that the unsupervised approach taken by \citet{Kounkel-Covey2019} resulted in many strings featuring kinematically hot values above $5\,\mathrm{km}\,  \mathrm{s}^{-1}$. Whilst the identification and characterisation of tidal tails in clusters is still a source of open debate, it is beyond the scope of this work which focuses its observations on the canonically recognised cluster cores. Furthermore, in the select few cases that targets were only identified as a member by a single study, these are typically stars belonging to the filamentary strings and outside the FoV of NGTS. Thus, we include members from \citet{Kounkel-Covey2019} \& \citet{Meingast2021} if they fall within the NGTS observing window, given the increased likelihood of membership in such a dense region (corroborated by their membership in multiple studies), however recommend some caution for more distant members when they are drawn from these references only.

\subsection{NGTS Observations}
\label{ssec ngts obs}

\begin{table*}
\centering
\caption{Tabular overview of cluster membership and observation dates covered in Section \ref{ssec ngts obs}}
\resizebox{\textwidth}{!}{
\begin{tabular}{|c|c|c|c|c|c|c|c|c|}
\hline  \hline
\textbf{Cluster} &
  \textbf{\begin{tabular}[c]{@{}c@{}}Cluster\\ centre $^{(*)}$\\ l (Deg)\end{tabular}} &
  \textbf{\begin{tabular}[c]{@{}c@{}}Cluster\\ centre $^{(*)}$ \\ b (Deg)\end{tabular}} &
  \textbf{\begin{tabular}[c]{@{}c@{}}$\textrm{Distance}^{(*)}$\\ (pc)\end{tabular}} &
  \textbf{\begin{tabular}[c]{@{}c@{}}\# Cands. \\$\textrm{full}^{(\mathrm{A})}$\end{tabular}} &
  \textbf{\begin{tabular}[c]{@{}c@{}}\# Membs. \\$\textrm{full}$\\($\geq70$\%)\end{tabular}} &
  \textbf{\begin{tabular}[c]{@{}c@{}}\# Membs. \\$\textrm{filt}$\\ ($\geq70$\%)\end{tabular}} &
  \textbf{\begin{tabular}[c]{@{}c@{}}\# Membs. \\ with NGTS\\ LC's\end{tabular}} &
  \textbf{\begin{tabular}[c]{@{}c@{}}Obs\\ dates\\ yyyy/mm/dd\end{tabular}} \\ \hline
Alessi 3      & 257.449 & $-15.064$ & 278.1 & 324 & 240  & 148   & 136  & 2021-11-03 : 2022-08-10   \\ 
Trumpler 10   & 262.866 & $0.582$   & 431.8 & 1831 & 1726 & 1048  & 465  & 2020-09-29 : 2021-07-05   \\ 
NGC 2451A     & 252.431 &$ -7.276$  & 192.5 & 1520 & 1506 & 331  & 307  & 2020-12-07 : 2021-08-17   \\ 
NGC 2451B     & 252.309 & $-6.856$  & 364   & 325 & 283  & 283   & 250  & 2020-12-07 : 2021-08-17   \\ 
Collinder 135 & 248.99  & $-11.201$ & 302.4 & 433 & 324  & 324   & 111  & 2020-09-29 : 2021-03-26  \\ 
NGC 2516      & 273.861 & $-15.873$ & 408.9 & 3800 & 3763 & 2551 & 1633 & 2019-11-30 : 2020-03-22  \\ 
IC 2391       & 270.386 & $-6.737$  & 151.3 & 743 & 739  & 328  & 129  & 2021-11-12 : 2022-09-30  \\ \hline
\end{tabular}
}
\label{tab: cl overview}
\vspace{1mm}
\parbox{\linewidth}{\small
\textit{NOTES.}\newline
$(*)$ -- All coordinates \& distances from \citet{Cantat-Gaudin2020}. $(A)$ -- Refer to appendix \ref{app cl memb tab} for a description of each column.}%

\end{table*}

NGTS comprises twelve 20 cm wide field telescopes at a remotely operated facility, situated at the ESO observatory in Paranal, Chile. Each telescope has a FoV of $2.8\degr$ and can achieve sub-milli-magnitude level photometric precision. In this work, we focus on seven young open clusters which were observed as part of the NGTS Clusters Survey. These clusters are: Alessi 3, Trumpler 10, Collinder 135, NGC 2451\,A\&B, NGC 2516 \& IC 2391. Details of each cluster and its observations are provided in the following subsections, with a summary of cluster membership and observations given in Table\,\ref{tab: cl overview} (see Appendix \ref{app cl memb tab} for a description of each column). We also note that, because the bulk cluster properties were typically slightly different in each literature study, for consistency, all distances and coordinates quoted in this work were obtained from \cite{2020A&A...633A..99C}.

\subsubsection{Light curve pre-processing}
Before extracting photometry several cuts were made to our target lists, namely;
\begin{enumerate}
    \item We require $\textrm{\textit{Gaia} G mag} < 17$. For targets fainter than this photometric precision is generally too low to measure rotation periods.
    
    \item The NGTS pipeline calculates a dilution value for each target star, given as the ratio of flux from contaminating sources over flux from the target star. Targets were cut if the dilution within the 3 pixel aperture used for photometric measurements exceeded a value of 0.5.
\end{enumerate}

As well as above, we required Teff $\leqslant 6500\,\mathrm{K}$ $(\mathrm{M} \lesssim 1.3\mathrm{M}_{\odot} )$ to ensure that only F, G, K and M type stars with convective envelopes amenable to rotation studies are considered. Stars outside this range have transitioned from convective envelopes to fully radiative. Photometric modulation in stars $(\mathrm{M} \gtrsim 1.3\mathrm{M}_{\odot} )$ is typically a result of pulsations and not surface activity\footnote{We note this is not the case for massive chemically peculiar stars.}. This filter was enforced post extraction however, so that separate work (not considered here) could be conducted on high mass stars. It is therefore, not reflected in the number of light curves extracted herein. Where available, we assign temperatures from Gaia's DR3 release and Gaia DR2 effective temperatures otherwise \citep{gaiadr2,babusiaux18_ext,gaiadr3}. Corresponding masses were calculated using the Gaia effective temperatures along with the MESA Isochrone \& Stellar Tracks (MIST) equivalent evolutionary point tracks \citep{MIST1,MIST2}. \par

Each cluster was observed independently by one or more telescopes for a baseline of $\sim$\,200 nights, with photometry extracted by the NGTS pipeline for cluster members that fell within the NGTS FoV. For all clusters, exposures were taken at 13s cadence and, as part of the NGTS pipeline, excess variance values were calculated for each cadence. For each night, the standard deviation of the excess variance statistic per frame was calculated. Where this value exceeded 0.01, indicating unstable photometric noise within the night (e.g. passing cloud, guiding issues, laser crossing events) that night's observations were removed, along with a $5\sigma$ cut across the entire light curve. For more information on the pipeline, we refer the reader to \cite{2018MNRAS.475.4476W}. After cleaning the light curve, the data was binned to a 5-minute cadence for our analysis. \par

\subsubsection{Alessi 3}
Alessi 3 is centered at $\mathrm{RA} = 7^h17^m06.00^s$ and $\mathrm{Dec} = -46\degr 08' 31.20''$ ($l =257.449\degr$, $b = -15.064\degr$ in galactic coordinates) at a distance of 278.1\,pc. All FGK single stars in this cluster should have reached the main sequence, converging onto the slow sequence and now be evolving via some characteristic mass-dependent period range. Alessi 3 was observed by NGTS from $13^{\mathrm{th}}$ November 2021 - $10^{\mathrm{th}}$ August 2022 (2021-11-13:2022-08-10). 170,058 exposures were obtained over a baseline of 214 days, with photometry on 136 members being extracted.

\begin{figure*}
    \centering
    \includegraphics[width = 0.96 \linewidth]{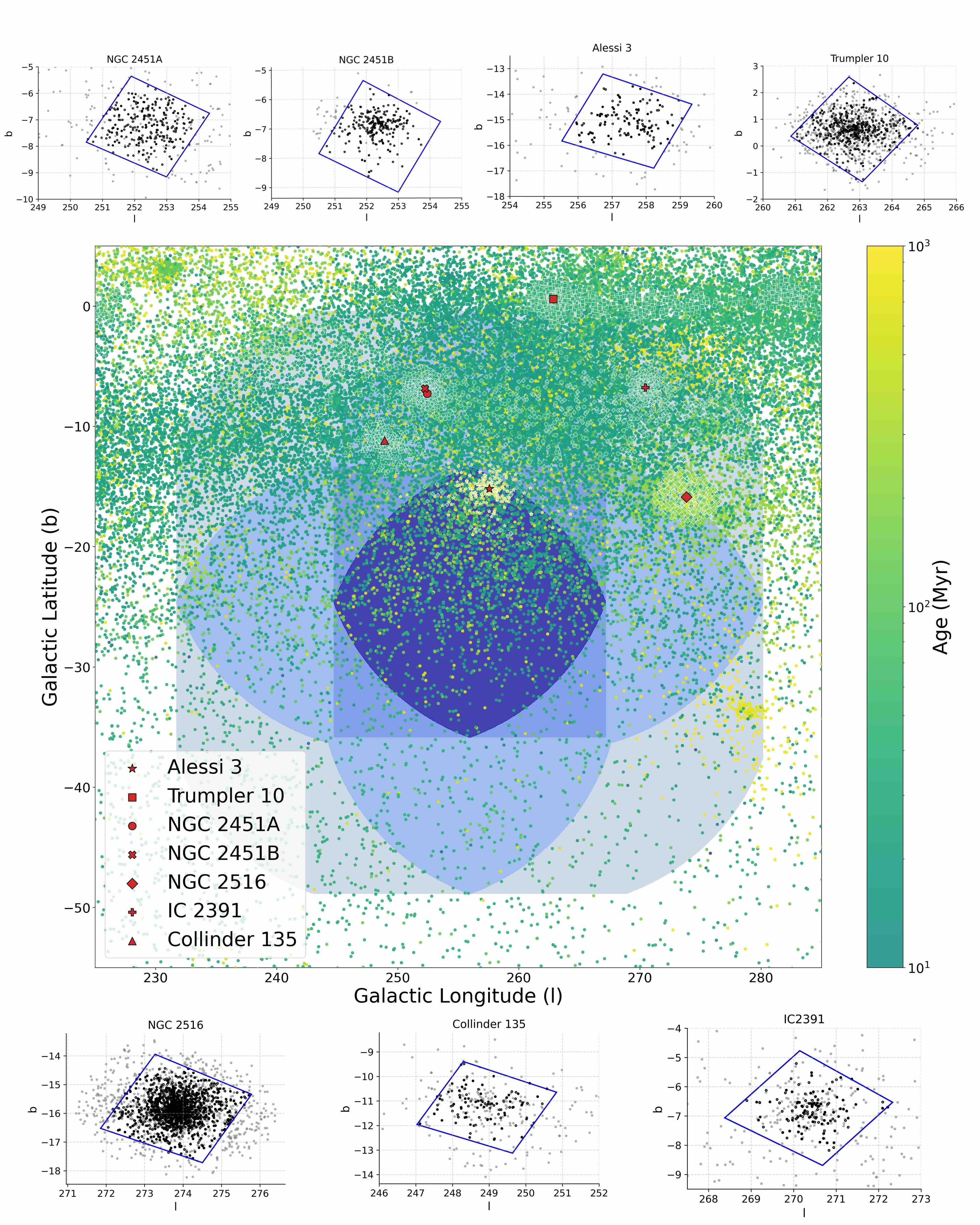}
    \caption{\textbf{Centre plot:} Positions of confirmed and candidate young stars (<1Gyr) with Gmag < 17 and distance < 500\,pc, highlighted by their ages, overplotted onto the PLATO LOPS2 field and its relative camera assembly overlaps. The central darkest blue corresponds to all 24 cameras, decreasing to six at the lightest outer edges. Clusters observed by NGTS are positioned on the plots according to their field centre outlined in section \ref{ssec ngts clusters}. Cluster members have been overplotted on the LOPS2 field with white circle outlines. Alessi 3 in particular falls almost entirely in the central region of the LOPS2 field, exposed to PLATO's full 24 camera array. Note whilst subplots focus on the cluster cores observed by NGTS, full membership including tidal tails identified by \citet{Kounkel-Covey2019} and \citet{Meingast2021} have been included in the central plot. \textbf{Outer subplots:} Spatial distributions of cluster members for the clusters observed in this study. Grey points indicate cluster members from our membership survey. Black outlined points are those with NGTS light curves, black filled points represent stars for which we returned rotation periods. The blue box corresponds to the NGTS FoV.}
    \label{fig YSC cl pos}
\end{figure*}

\subsubsection{Trumpler 10}
Trumpler 10 is centered at $\mathrm{RA} = 8^h47^m46.320^s$ and $\mathrm{Dec} = -42\degr 33' 57.60''$ ($l =262.866\degr$, $b = 0.582\degr$ in galactic coordinates) at a distance of 431.8\,pc. Trumpler 10 was observed by NGTS from 2020-09-29:2021-07-05. 206,857 exposures were obtained over a baseline of 218 days. Photometry for 465 stars was extracted. It resides at the upper corner of the LOPS2 field, and is only partially inside the field itself. However, given that a sizable portion of the central cluster is observable, it has been included as a part of this study.

\subsubsection{NGC 2451}
Although there is some disagreement on who discovered NGC 2451, credit for its formal identification is attributed to John Herschel \citep{ngc2451b_ext}. Initially, it was identified as one dense cluster in the constellation Puppis, however more recent studies have confirmed that NGC 2451 is actually two distinct line-of-sight clusters, which despite being spatially independent, are labeled NGC 2451 A \& B. \par

\noindent NGC 2451\,A is the closer of the two, centered at $\mathrm{RA} = 7^h42^m56.64^s$ and $\mathrm{Dec} = -38\degr 15' 50.40''$ ($l =252.431\degr$, $b = -7.276\degr$ in galactic coordinates) at a distance of 192.5\,pc. It is one of the closest known young open clusters to us, making it an excellent candidate for photometric study of stellar evolution. \par
\vspace{2mm}

\noindent Whilst NGC 2451\,B lies almost directly behind NGC 2451 A on the sky, centered at $\mathrm{RA} = 7^h44^m30.72^s$ and $\mathrm{Dec} = -37\degr 57' 14.40''$ ($l =252.309\degr$, $b = -6.856\degr$ in galactic coordinates), the two clusters are actually widely separated with NGC 2451\,B almost twice as far as A, at 364\,pc. NGC 2451\,B has been studied substantially less than NGC 2451\,A, largely due to previous difficulties of distinguishing cluster members from foreground member stars of NGC 2451\,A. This also makes photometric study harder due to increased blending in such a dense environment, requiring precision and detrending that historically has not been achievable.  \par

NGTS observed both NGC 2451 clusters simultaneously from 2020-12-07:2021-08-17, obtaining 166,480 exposures over that period. Photometry on 577 targets were extracted by NGTS, with 307 targets in NGC 2451\,A and 250 in NGC 2451\,B.

\subsubsection{Collinder 135}
Collinder 135 is centered at $\mathrm{RA} = 7^h17^m26.88^s$ and $\mathrm{Dec} = -37\degr02' 38.40''$ ($l =248.99\degr$, $b = -11.201\degr$ in galactic coordinates) at a distance of 302.4\,pc. Collinder 135 was observed by NGTS from 2020-09-29:2021-03-26. In total, NGTS captured 185,885 exposures over a baseline of $~170$ days, with photometry on 111 members within the FoV meeting our threshold for extraction.  

\subsubsection{NGC 2516}
NGC 2516 is a large open cluster, centered at $\mathrm{RA} = 07^h 58^m 06.48^s$ and $\mathrm{Dec} = -60\degr 48' 00.00'' $ ($l =273.861\degr$, $b = -15.873\degr$ in galactic coordinates) at a distance of 408.9\,pc. Recent observations by \cite{2021AJ....162..197B} identified an extended halo, extending the cluster over 500\,pc in size. NGTS observed the central core of the cluster, taking a total of 167,483 exposures from 2019-11-30:2020-03-22. Light curves for 1633 targets met our extraction criteria. 

\subsubsection{IC 2391}

IC 2391 is centered at $\mathrm{RA} = 08^h 41^m 10.00^s$ and $\mathrm{Dec} = -52\degr 59' 27.60'' $ ($l =270.386\degr$, $b = -6.737\degr$ in galactic coordinates) at a distance of 151.3\,pc. IC 2391 was observed by NGTS from 2021-11-12:2022-09-30. In total, NGTS captured 202,279 exposures over a baseline of $~170$ days, with photometry on 129 members within the FoV meeting our threshold for extraction.  

\subsection{Cluster ages}
\label{ssec ages}
\begin{figure}
    \centering
    \includegraphics[width = \linewidth]{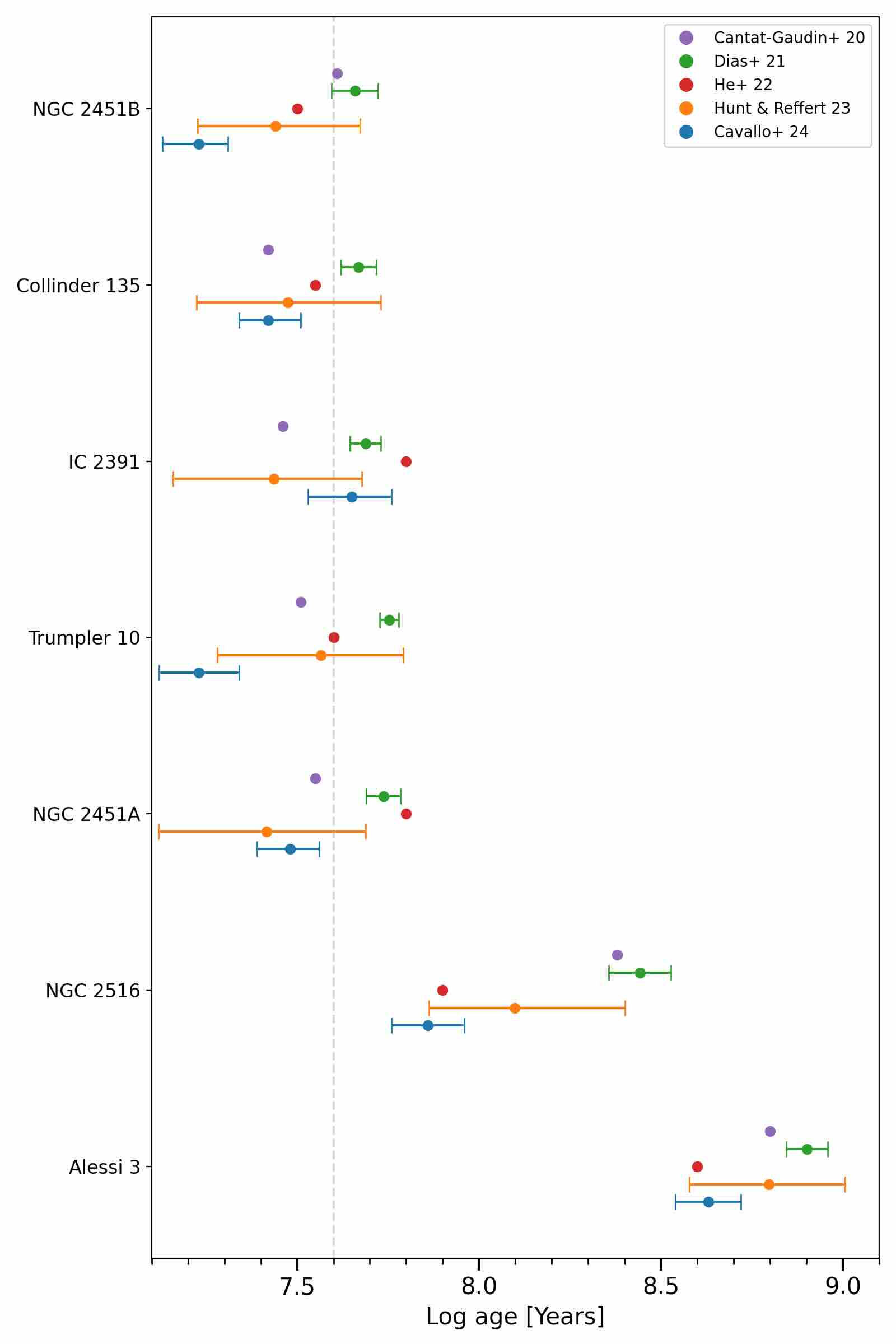}
    \caption{Literature ages from studies post Gaia DR2 data release, that attained age estimates for target clusters in this work. The grey dashed line highlights the age of the ZAMS for a sun like star.}
    \label{fig: lit ages}
\end{figure}

Following the process in \cite{2024ApJ...962...16D} we conducted an independent literature review of cluster ages. The selection criteria for the literature review followed that of our membership criteria, in that only studies post Gaia DR2 were considered. This ensures that estimates are based on more precise observational data. Additionally, each study derived ages for either all, or the majority of the clusters in this work. Any systematic differences in age estimates between studies should remain consistent across clusters, allowing for a more meaningful age comparison.\par

Figure \ref{fig: lit ages} shows the resulting age estimates for clusters from five different studies \citep{2020A&A...640A...1C,2021MNRAS.504..356D,2022ApJS..262....7H,2023A&A...673A.114H,2024AJ....167...12C}. Of the five studies considered, \citeauthor{2021MNRAS.504..356D} \& \citeauthor{2022ApJS..262....7H} explicitly derived cluster ages through isochrone fitting, whereas the remaining three employed machine learning approaches based on neural networks (NNs). Specifically, \citeauthor{2020A&A...640A...1C} and \citeauthor{2024AJ....167...12C} utilised artificial neural networks (ANNs), while \citeauthor{2023A&A...673A.114H} adopted a convolutional neural network (CNN) framework. Each of these methodologies are widely used in large-scale studies of stellar clusters, where automation and uniform processing are essential.\par

However, phenomena such as binaries, the presence of blue stragglers, differential extinction, rotation and the intrinsic spread of low-mass stellar populations can all affect the morphology of a cluster's colour-magnitude diagram (CMD) \citep{bluestragcmd,cmdbroadening, rotationcmd}. When training NNs on synthetic CMDs, it is crucial that these processes are accurately modeled, however, doing so is computationally expensive and complex, particularly in large automated studies and is therefore often neglected or oversimplified. Conversely, when using observational CMDs, many variations are required to properly constrain the full distribution of possible CMD morphologies at a given age and usually require auxiliary parameter indicators such as age estimates, reddening values etc which are not always readily available. As a result, both synthetic and observational training approaches can suffer from systematic biases in age estimation due to algorithmic sensitivity to CMD perturbations caused by the processes outlined above, among others. This sensitivity likely contributes to the wide dispersion in age estimates observed across the literature. Notably, this spread is evident not only between studies employing different methods but also among those using the same technique, further highlighting the role of systematic uncertainties on output parameters.

For all clusters bar Alessi 3 and NGC 2516, ages are constrained between $\sim$\,20\,--\,70\,$\textrm{Myr}$, with the grey dashed line in Figure \ref{fig: lit ages} marking the $40$\,Myr point at which solar mass stars would reach the ZAMS \citep{MIST2,2017A&A...599A..49K}. NGC 2451\,B has the lowest average age estimate of the clusters in this work, with \citeauthor{2024AJ....167...12C} citing an age $<20$\,Myr. Interestingly, \citeauthor{2024AJ....167...12C} reported that their estimates for young clusters ($<100$\,Myr) are typically systematically older than that of \citep{2021MNRAS.504..356D, Cantat-Gaudin2020}. Their analysis of clusters examined in this work however show the opposite and provide systematically lower age estimates than the aforementioned studies. This discrepancy shows the often substantial and conflicting uncertainties on age predictions in young clusters, which can have important consequences when interpreting the clusters evolutionary state.
For example, \citeauthor{2024AJ....167...12C} provides an age estimate of $~17$\,Myr for Trumpler 10, markedly lower than other studies which have it at least double that age. Whilst a difference of $\sim15$\,Myr is negligible for mature main sequence stars ($>1$\,Gyr), at younger ages it can be the difference between undergoing fusion or not, marking substantial differences in the stars evolutionary state. Given the accuracy required at such young ages for meaningful investigation coupled with the current spread of age estimates in the literature, we refrain from adopting specific ages from isochronal fitting for the clusters in this work. Instead, we note the recent photometric study by \citet{2023AJ....166...14B} of the young open cluster Alpha Persei (herein referred to as \alper). With a well constrained age of $79^{+1.5}_{-2.3}$\,Myr \citep{2022A&A...664A..70G}, \alper currently forms the canonical lower age anchor for gyrochronology. This age estimate is older than those in the literature for the younger clusters in this work, which is corroborated later in section \ref{sec disc} by the rotation profiles of all five clusters (i.e. they are less evolved and exist at or below the sequence of \alper). Thus we conclude that whilst adopting specific ages is unreliable, given the substantial and often conflicting uncertainties of age estimates in young clusters from isochronal fitting, all clusters other than Alessi 3 and NGC 2516 are $\lesssim80$\,Myr with NGC 2451\,B appearing comparatively younger than the other clusters. 

\section{Rotation period characterisation}
\label{sec measuring prot}

\subsection{Period detection}
\label{ssec method}

Due to the number of stars processed in this work, an automated pipeline was created in order to analyse, filter and report results for each cluster in a homogeneous manner.\par
To start, the Lomb-Scargle (LS) periodogram was calculated for a given cluster member using \texttt{ASTROPY} \citep{astropy:2022} between $0.1-50$\,d. For a detailed overview of the LS periodogram, we encourage the reader to review \citet{2018ApJS..236...16V}. For the purposes of this work, it suffices to summarise that LS fits a series of sinusoidal models to the dataset, using the least squares fitting approach to assign the relative power of each model. LS is a popular method for detecting and characterizing periodicity in astrophysics and has been widely used to identify stellar rotation periods in the past. 

Following the process of \citet{2020MNRAS.492.1008G} and \citet{2023MNRAS.523..169S}, we conduct an initial check for moon contamination. Unlike these studies, which limit their checks to only the faintest stars, this step was performed for every star in our analysis. The justification for this is that whilst identifying moon contamination signals in young stars is relatively straight forward, LS will often also identify alias peaks for this contamination at both $\sim 10$\,d \& $\sim15$\,d. For clusters older than $\sim 200$\,Myr, these alias peaks could represent real rotation signals and thus for such clusters its important to identify possible contamination whilst vetting these periods. Furthermore, we find that in the process of detrending moon contamination the relative power of remaining peaks can be affected, meaning this step is important in order to accurately report on rotation. The computational expense to perform this check is negligible and therefore it is included for each cluster to ensure an accurate, homogeneous approach to our analysis. The bounds for alias periods are defined according to \citet{2023MNRAS.523..169S}, where a period was classed as an alias if it fell within bounds set by
\begin{equation}
    \textrm{Bounds} = \left (P_{\textrm{obs}} - 2\frac{P_{\textrm{obs}}}{\textrm{baseline}}, P_{\textrm{obs}} + 2\frac{P_{\textrm{obs}}}{\textrm{baseline}} \right)
\end{equation}

\noindent
where the observed period $P_{\textrm{obs}}$ is defined as
\begin{equation}
    P_{\textrm{obs}} = \left( \frac{1}{P_{\textrm{true}}} + n \right)^{-1}
\end{equation}

\noindent
for n in $\pm$\,[1, 2, 3, 4].

\par
Any star with a primary periodogram peak between $27\,\mathrm{d} < \mathrm{P} < 30\,\mathrm{d}$, or that had a peak in this range which was greater than half the power of the primary peak, was considered to be contaminated by the moon. The light curve was subsequently phase folded on the identified contaminant period which was removed in two stages, by applying a Savitsky-Golay filter followed by a convolution. Once the data had been detrended, a new LS periodogram was calculated to ensure the contaminated signal had been effectively removed. At this stage, the periodogram results were also filtered to remove systematic peaks resulting from the diurnal nature of ground based observation. For almost all short period targets these peaks ($\sim 1\,\mathrm{d}$ and the aliases for $\mathrm{n} \leq 4$) were among the strongest peaks in the periodogram. Once these peaks had been filtered, the highest power remaining peak was assigned as the provisional rotation period, as well as secondary and tertiary period estimates\footnote{False alarm probabilities for all period estimates were essentially zero, with a maximum value of $10^{-6}$ and a median of $10^{-136}$.}. Any period that fell within the alias bounds for $\mathrm{n} = \pm 1,2$ of the lunar period were not removed but flagged as potentially contaminated, subject to inspection.

\subsection{Period vetting}
\label{ssec vetting}

Once a preliminary period estimate was identified, the data was phase-folded on both the primary and secondary period estimates separately. These two periods were visually inspected alongside the detrended periodogram in the form of a diagnostic plot that contained the light curve, target star metadata and both the phase folded plots alongside the periodogram. Where clear periodicity in either of the phase folds was present, or a clear peak in the periodogram, the target was passed forwards as a potential rotation signal.\par

\begin{figure*}
    \centering
    \includegraphics[width=1\linewidth]{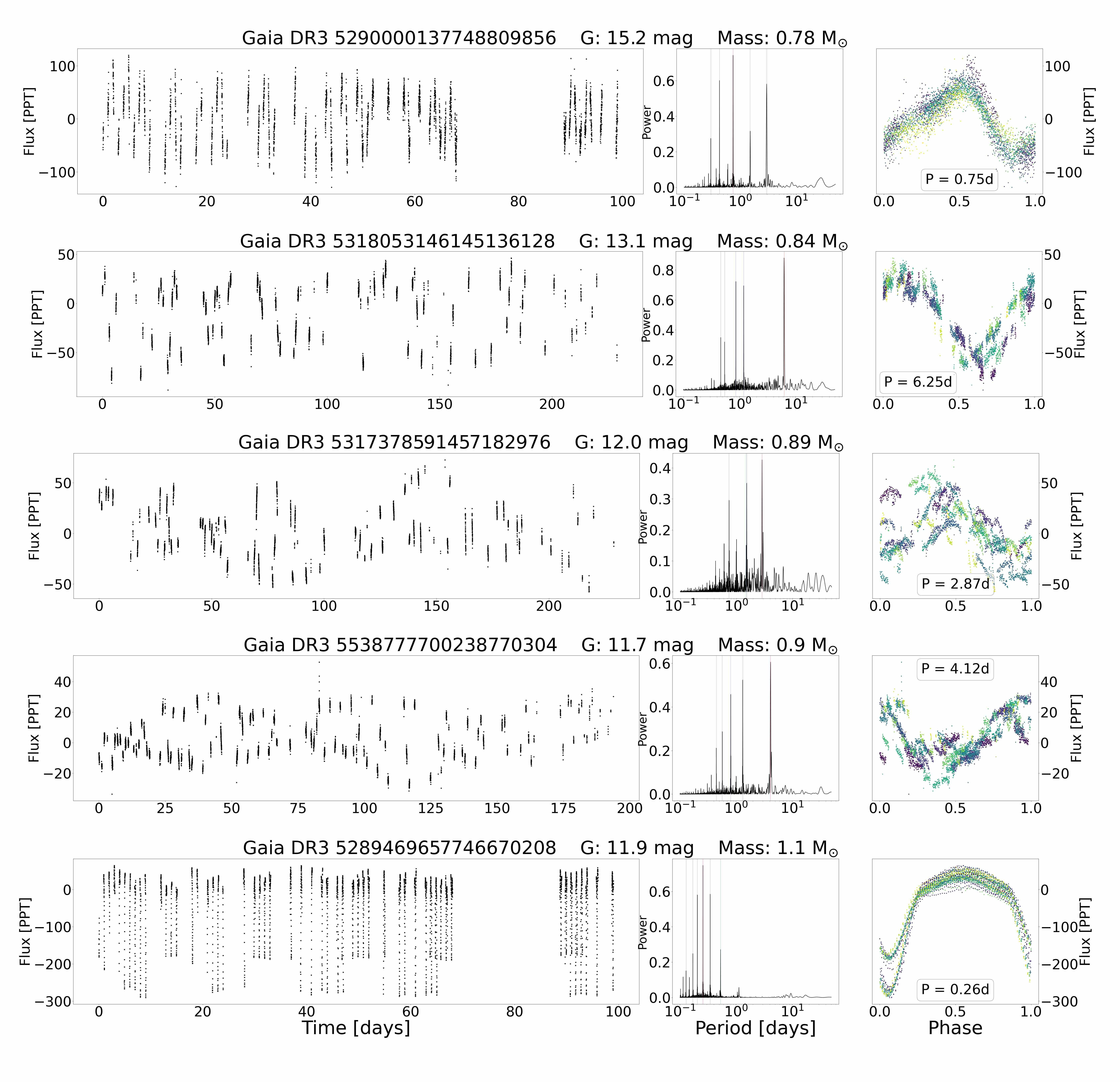}
    \caption{Selection of different NGTS light curves and diagnostic plots from our period characterisation pipeline. Left hand panels: NGTS light curves, post any detrending and normalised to parts per thousand. Middle panels:Lomb-Scargle periodogram. Light blue vertical lines represent periods identified as alias or harmonic periods of the adopted period, which is identified by a red vertical line. Right hand column: light curve phase folded on the adopted period. The phase folded data cycles through a colour map in time, from dark blue to yellow to identify potential amplitude \& phase changes that can help constrain the correct period (this is clear in row two, where plotting as a single colour would appear to show the phase folded data having a large spread, which is actually a result of changing amplitude). The final row shows a binary system, where Lomb-Scargle has identified the orbital period.}
    \label{fig lc examples}
\end{figure*}

We utilised the Lomb-Scargle best fit model function within the \texttt{ASTROPY} package \citep{astropy:2022} package which, at a fixed frequency, uses a linear least-squares approach to determine the best fitting amplitude, phase and offset for the sinusoidal model. Primary, secondary and tertiary period estimates were passed into the Lomb-Scargle model function to produce a best fit model for each period which was over-plotted on the detrended light curve. Alongside visual inspection of the phase folded light curve, comparison of the best fit models was typically sufficient to identify the true period. However where available, extra checks against TESS data were performed for each target, opting to utilise the \texttt{LIGHTKURVE} package\footnote{Lightkurve utilises \texttt{ASTROQUERY} \citet{astroquery} in order to access data stored on the Mikulski Archive for Space Telescopes (MAST).} \citep{lightkurve} to download light curves from the Cluster Difference Imaging Photometric Survey (CDIPS) \citep{bouma2019cluster} where possible. Although TESS has a pixel scale of $\sim21^"$ \citep{tess}, much larger than that of NGTS at $\sim5^"$ \citep{2018MNRAS.475.4476W} and is thus more susceptible to contamination, they often provided further confirmation of our rotation period estimates. These checks were especially beneficial in cases where the initial primary peak was around one day (or an alias) and thus subsequently removed. Because TESS is not susceptible to the same ground based systematic signals, where agreement was found between NGTS and TESS data on such periods, they were reinserted as primary period estimates and rechecked following the same process as described above. Figure \ref{fig lc examples} show some examples of detrended NGTS light curves, alongside the Lomb-Scargle periodogram and light curves phase-folded on the period adopted by this work. 

\subsection{Identifying higher order systems}
\label{ssec binaries}
\begin{figure}
    \centering
    \includegraphics[width=\linewidth]{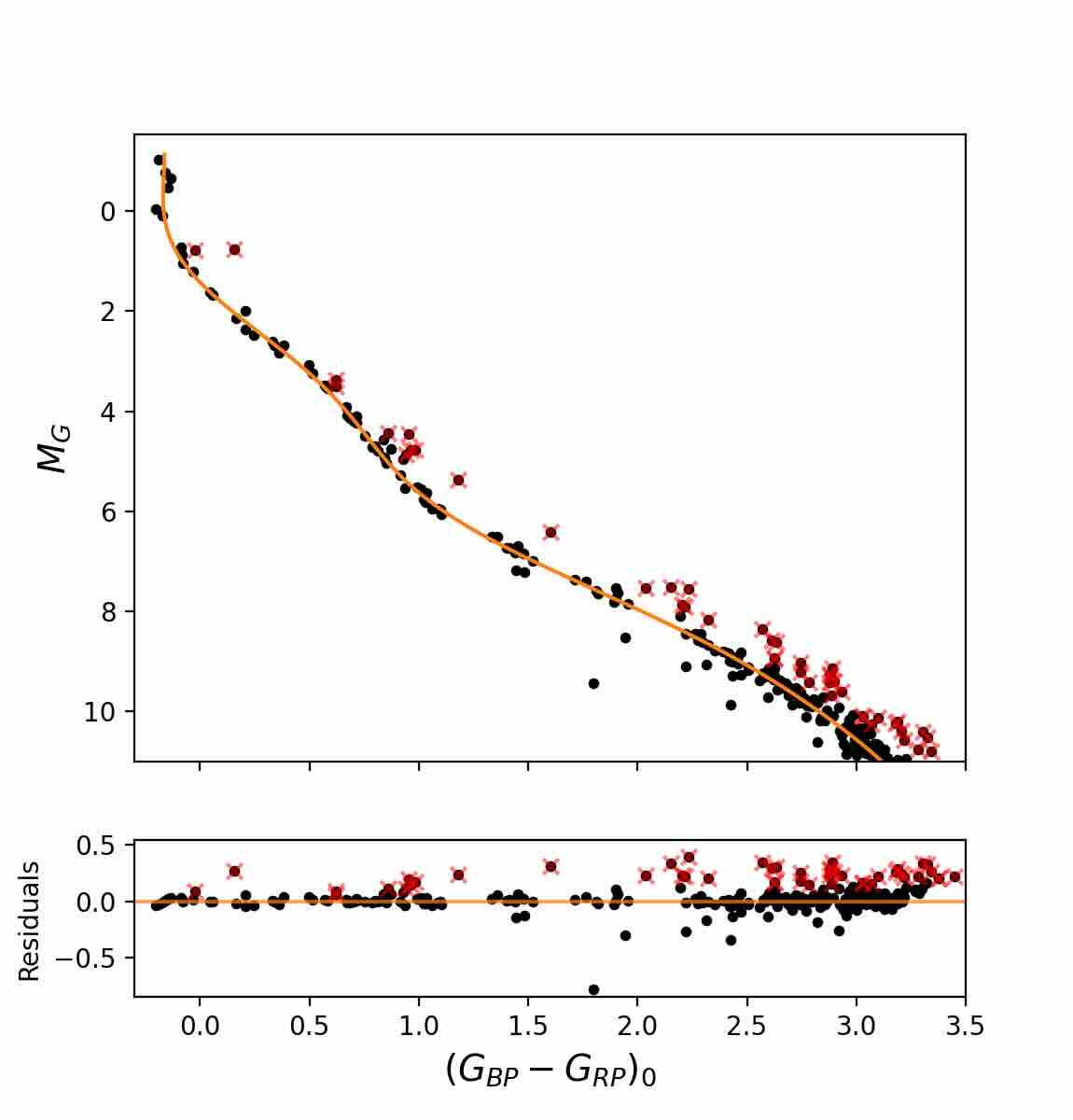}
    \caption{Top: Colour magnitude diagram of absolute Gaia G magnitude versus dereddened Gaia BP-RP colour for NGC 2451\,A. The orange line shows the maximum a posteriori Gaussian process (GP) fit to the cluster sequence, with red crosses highlighting stars that lie outside of the threshold outlined in section \ref{ssec binaries} and represent likely higher order systems. Bottom: Residuals of the fit, following \citet{2020MNRAS.492.1008G}, where the `residual' is calculated as the smallest linear distance from the best fit single star sequence.}
    \label{fig cmd}
\end{figure}

If one wants to investigate stellar evolution, it is important to first identify higher-order systems (e.g. binary or triple star systems). The presence of a companion star will induce complex and exotic interactions, such as gravitational tidal forces and potential mass transfer. These processes are not experienced by isolated single stars and thus single stars and higher order systems undergo vastly different evolution \citep[e.g.][]{2010IAUS..262...44H,2025CoSka..55c..21B}. Thus, in order to probe the intrinsic spin down of a single star due to its own angular momentum evolution (the underlying concept of gyrochronology) it is important to make the distinction between single and higher order star systems. Understanding the distribution of these systems over a prolonged period of their youth can also give a better understanding of their independent evolutionary pathways.\par

To identify higher-order star systems, we utilise two methods. Primarily, higher-order systems are identified by iteratively fitting CMD's using a Gaussian process (GP) and rejecting outliers that did not meet a threshold $\sigma$ level, following the approach adopted in \cite{2020MNRAS.492.1008G}.
We create a CMD in absolute Gaia magnitude $\mathrm{M}_\mathrm{G}$ vs. dereddened colour $(\mathrm{G}_{\mathrm{BP}}-\mathrm{G}_{\mathrm{RP}})_0$, using an extinction coefficient of $3.1$ \citep{ext_rv1, ext_rv2} and cluster specfic colour excess values. Extinction values were calculated and the data dereddened using the extinction ratios for Gaia DR3 from the SVO filter service \citep{svo1,svo2,svo3}. Specific colour excess values, and appropriate references are listed in table \ref{tab ext}. \par

We iteratively fit a GP to the entire cluster sequence, choosing to use the squared-exponential (SE) kernel in \texttt{TINYGP} \citep{tinygp}. The SE kernel was chosen over the Matern-3/2 due to the Matern's sensitivity to noisy data \citep{matern}, thus increasing the chance of missing binary stars when modelling smaller clusters with higher levels of scatter. The SE produces a smoother interpolation which, although it is less optimised to the turn-off region of the CMD than the Matern, produces an overall better fit. Prior to rejecting outliers, the hyperparameters of the kernel, namely amplitude and scale factor, were optimised by training the GP on the observational data. The optimised kernel parameters were then used to fit the GP to the data and reject outliers at a level of $2\sigma$. The reason behind rejecting outliers at $2\sigma$ rather than $3\sigma$ was that in cases of smaller clusters, outliers would have a larger effect on our rolling sigma value, a result of fewer stars in each window. This resulted in the rolling $\sigma$ threshold being inflated in lower density regions of the CMD and not identifying all clear outliers when visually inspected. The decision was therefore taken to iteratively cut outliers at a harsher $2\sigma$ level until convergence on the single star sequence (typically within $\sim 7$ iterations). A final rejection of higher-order systems was performed on the whole sequence at a level of $3\sigma$ from the converged model. Figure \ref{fig cmd} shows the CMD in $\mathrm{M_{G}}$ vs. $(\mathrm{G}_{\mathrm{BP}}-G_{\mathrm{RP}})_0$ for NGC 2451\,A. The final converged slow sequence model is overplotted in orange and stars that failed the final $3\sigma$ rejection test on the converged sequence are highlighted with red crosses. To ensure consistency with our results, we employ this technique to identify higher order systems in all clusters. \par

\begin{table}
\centering
\caption{Extinction values used for clusters in this study in order to deredden data for Colour magnitude diagrams, with corresponding reference for each value.}
\begin{tabular}{|lll|}
\hline \hline
\multicolumn{1}{|l|}{\textbf{Cluster}} & \multicolumn{1}{l|}{\textbf{E(B-V)}} & \textbf{ref} \\ \hline
NGC 2451\,B & 0.05 & \protect\cite{ngc2451b_ext} \\
Collinder 135 & 0.113* & \protect\cite{He22_ext} \\
IC 2391 & 0.03 & \protect\cite{babusiaux18_ext} \\
Trumpler 10 & 0.056 & \protect\cite{babusiaux18_ext} \\
NGC 2451\,A & 0.000 & \protect\cite{babusiaux18_ext} \\
NGC 2516 & 0.071 & \protect\cite{babusiaux18_ext} \\
Alessi 3 & 0.14 & \protect\cite{zerjal_ext} \\
\hline
\end{tabular}
\label{tab ext}

\parbox{\linewidth}{\small
\textit{Note.} $^{*}$ Actual value used in analysis was extinction value quoted by \citeauthor{He22_ext}. E(B-V) calculated for completeness.
}
\end{table}

As well as fitting CMDs to identify photometric binaries, we complemented our analysis by identifying astrometric binaries using Gaia's renormalised unit weight error (RUWE) threshold for a single star solution of RUWE < 1.4 (see e.g. \citealt{ruwe})\footnote{We note some studies choose to use a RUWE threshold for binarity of RUWE < 1.2. We include this filter in our results file for reference (see Table \ref{tab: results table} in the appendix) but choose a threshold of 1.4 during analysis.}. The combination of both methodologies ensures higher-order systems are identified as effectively as possible, with a star identified as higher order if it is flagged by either method.

\newpage
~
\newpage
~
\newpage

\subsection{Comparison to literature rotation periods}
\label{ssec lit comp}

\begin{figure*}
    \centering
    \includegraphics[width=1\linewidth]
{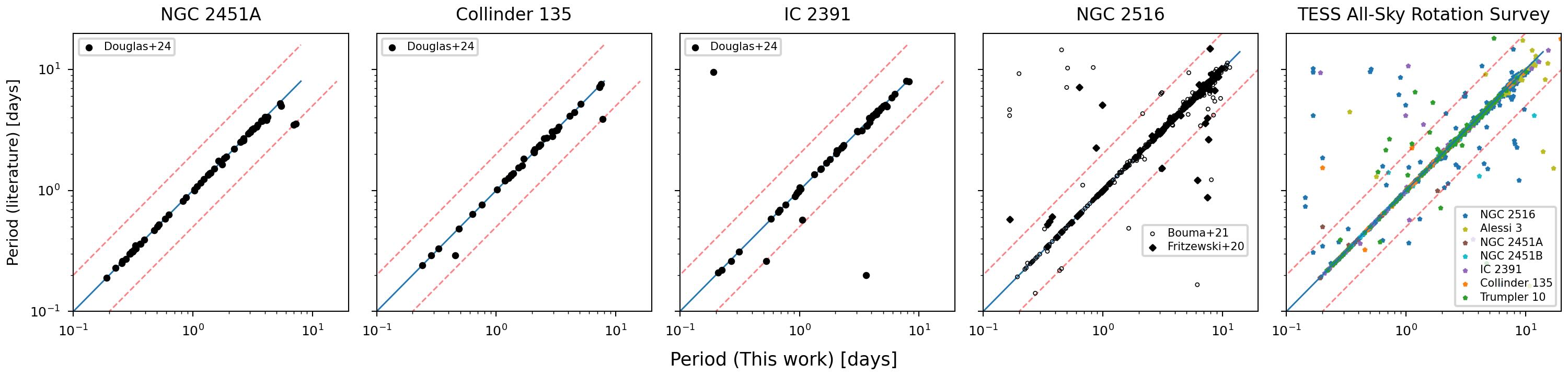}
    \caption{Period comparison for all clusters with existing literature studies. Left to right: panels one-to-three compare periods from this work with \citet{2024ApJ...962...16D}. Panel four compares to \citet[][open circles]{2021AJ....162..197B} \& \citet[][diamonds]{2020A&A...641A..51F}. Panel five compares this work to the TESS All-Sky Rotation survey \citep{2026arXiv260305586B}. For all panels, the blue line represents the 1:1 Period comparison line, with dashed red lines highlighting double or half-period harmonics.}
    \label{fig period comp}
\end{figure*}

After vetting, we compared the adopted rotation periods for stars in common with previous cluster-specific rotation studies in the literature to verify our method and results.
\cite{2024ApJ...962...16D} previously studied NGC 2451\,A, IC 2391 \& Collinder 135, finding good agreement between their rotation sequences and the wider literature and, as such, provides a good benchmark for comparison.
NGC 2516 has been extensively studied \citep{2007MNRAS.377..741I,2020A&A...641A..51F, 2021AJ....162..197B} with \cite{2021AJ....162..197B} recently identifying an extended halo spanning 500\,pc, although our NGTS observations focus on the dense core of the cluster (see section \ref{ssec ngts obs}). Observations by NGTS, along with \citeauthor{2007MNRAS.377..741I}, \citeauthor{2021AJ....162..197B} \& \citeauthor{2020A&A...641A..51F}, provide a comprehensive rotation period distribution for NGC 2516.\par

During the preparation of this work, \citet{2026arXiv260305586B} published a substantial all-sky rotation survey (referred to herein as TARS) for bright ($\mathrm{T} < 16$) targets within 500\,pc, reporting 1,046,317 rotation periods in total. Whilst \citeauthor{2026arXiv260305586B} is not a dedicated cluster survey, with 1000 stars commonly reported between this work and TARS, it provides another opportunity to compare this work to the wider literature.\par

Figure \ref{fig period comp} compares our rotation periods to those of the aforementioned studies for stars in common. Overall, this work is in very good agreement with the literature. Disagreements were typically the result of period harmonics, with the literature period generally being half the period adopted by this work. Based on comparisons of the high precision, long baseline NGTS observations with all available \textit{TESS} data from CDIPS \& Quick Look Pipeline (QLP) \citep{2020RNAAS...4..204H,2020RNAAS...4..206H} for each discrepancy we conclude that, shorter baseline observations by the literature were particuarly sensitive to identifying double dip modulation patterns on half the true period. In almost all cases we find the NGTS reported period here is the true period. Specific outliers within each cluster, including cases where we determined the literature period to be correct, are discussed in appendix \ref{app pcomp}. Refer also to Appendix Figure's \ref{app diagnostic}\,\&\,\ref{app tars} for examples of period discrepancies between this work and the literature values.

\section{Discussion}
\label{sec disc}

\subsection{Rotation period distributions}
\label{ssec period dists}

\begin{figure*}
    \centering
    \includegraphics[width =\linewidth]
    {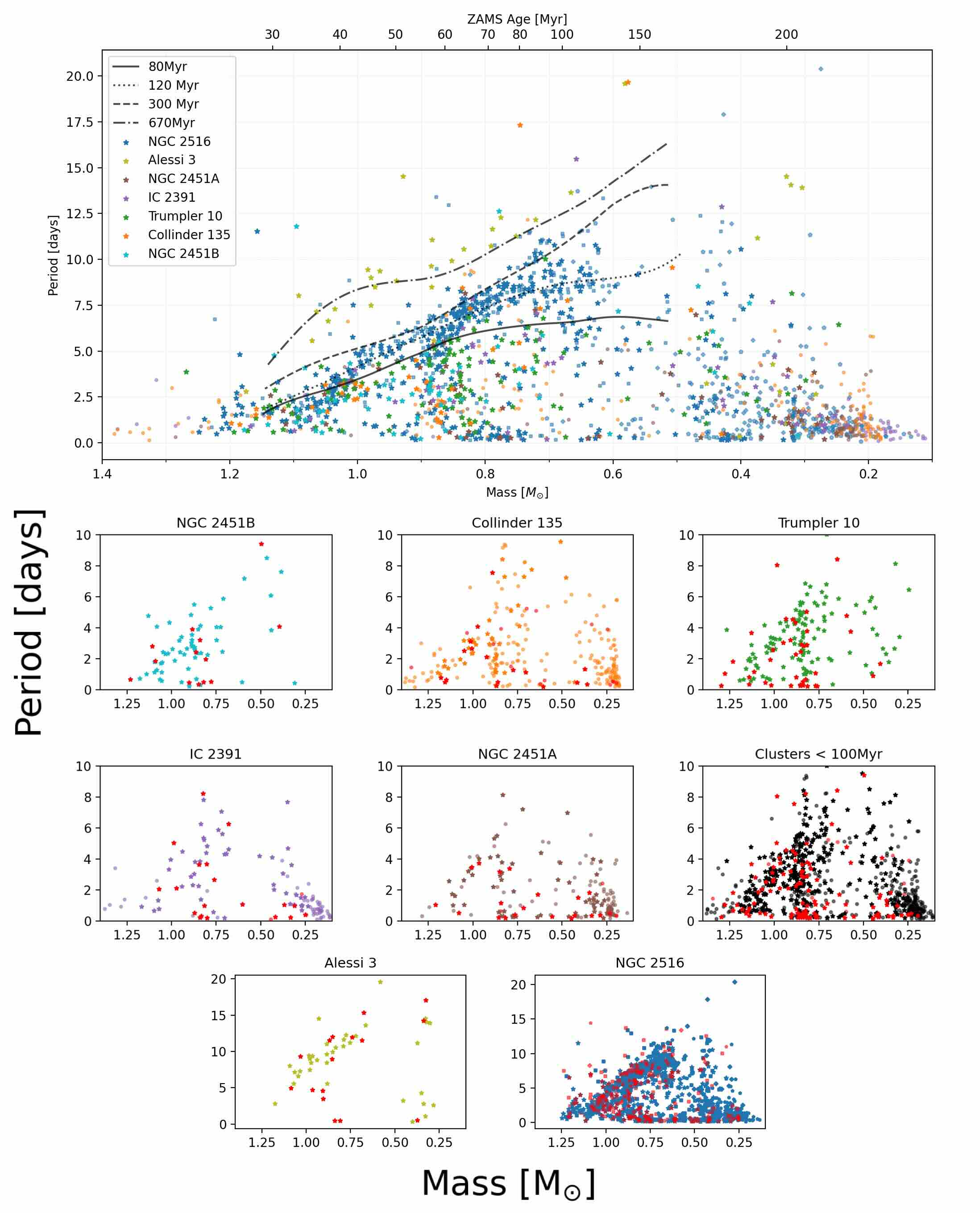}
    \caption{Top plot: Full single star period distribution for each cluster in this work. Circle, square and diamond markers are from 
    \citet{2024ApJ...962...16D}, \citet{2021AJ....162..197B} and \citet{2020A&A...641A..51F},
    respectively.
    Gyrochronal models from \citet{2023ApJ...947L...3B} have also been overplotted at key ages.
    Subplots: Individual rotation period distributions for each cluster. Markers follow the same notation as above. Red points represent identified higher order systems.}
    \label{fig period dist}
\end{figure*}

Across the seven clusters observed in the LOPS2 field we report 1063 rotation periods in total, with 479 being newly analysed as part of cluster specific rotational studies, however a majority of these targets were subsequently included in the recently published TESS All-sky Rotation Survey \citep{2026arXiv260305586B}. 63 periods in total are previously unreported. Of the 1063 stars with adopted rotation periods; 72 are in \textit{NGC 2451\,B}, 59 in \textit{Collinder 135}, 160 in \textit{Trumpler 10}, 69 in \textit{IC 2391}, 75 from \textit{NGC 2451\,A}, 52 in \textit{Alessi 3} and 576 from \textit{NGC 2516}; with the distributions for NGC 2451\,B, Trumpler 10 \& Alessi 3 being the first substantial, cluster specific rotation distribution results to date. 285 of the 1063 systems were also identified as likely binary or higher order systems (refer to section \ref{ssec binaries} for an overview of the process).\par

Figure \ref{fig period dist} outlines the rotation period distributions, including any identified higher order systems for all seven clusters. As outlined in section \ref{ssec ages}, five of the seven clusters have similar literature age estimates, all sub $80$\,Myr. In order to create the most complete rotation sequences possible, we compare the periods attained in this work to the wider literature and supplement our distributions with unique periods from previous studies. Figure \ref{fig period dist} shows the full period distributions of all clusters studied in this work, including literature periods to help extend down to lower masses. Until recently, gyrochronology studies have focused on clusters $\gtrsim$\,100\,Myr, arguing that they have had sufficient time to spin down onto their characteristic, mass-dependent rotation sequences. \cite{2023ApJ...947L...3B} recently provided a new lower age anchor for gyrochronal models, using a study of the $\sim$\,80\,Myr $\alpha$ Persei (\alper) cluster by \cite{2023AJ....166...14B} who noted the convergence of stars for masses $\gtrsim 0.8M_{\sun}$. The central image of figure \ref{fig period dist} shows period distributions for single star rotators, with gyrochronology models using \texttt{GYROINTERP} \citep{2023ApJ...947L...3B} providing a wider context for the mass-dependent evolution of rotation periods.\par
The rotation period distribution for Alessi 3 in this work is the first dedicated rotation distribution for the cluster and indicates an age roughly co-eval with the 670\,Myr gyrochrone that was attained by modeling Praesepe. A wide range of literature age estimates for Alessi 3 exist from $~400-800$\,Myr (see figure \ref{fig: lit ages}), with this work suggesting that the true age lies roughly in the middle of the literature range. The rotation distribution of Alessi 3 and Praesepe, along with an age determination for Alessi 3 are discussed in more detail in section \ref{ssec dga}.\par

\subsection{Extending the gyrochronal lower age limit}
\label{ssec lowering gyro limit}

Through the PMS, stars spin up as they contract to conserve angular momentum. An initial phase of disk-locking, that temporarily prevents spin up, leads to a spread of rotation periods by the age of the ZAMS. Most stars still cluster around a characteristic mass-dependent rotation period known as the slow sequence, while those with shorter periods, lying below this sequence in rotation-mass space, are referred to as fast rotators. The extent to which a clusters rotation sequence at a given age is identifiable is thus dependent on the extent to which fast rotators have converged onto this sequence once they have arrived at the main-sequence. \par

Whilst all clusters in this work have age estimates from processes such as isochronal fitting, at younger ages these estimates have large uncertainty (see Figure \ref{fig: lit ages}). \cite{2023MNRAS.523..802J} provided age estimates for multiple clusters spanning the entire range of stellar youth by modeling their photospheric lithium abundance, a complementary and arguably more precise method for age-dating young clusters. Four of the clusters in this work were included in their analysis (NGC 2451 A\&B, IC 2391 and NGC 2516). NGC 2451\,B is the youngest cluster with an age estimate of $43\pm 3$\,Myr and therefore provides an ideal opportunity to explore whether it is possible to extend gyrochronal models to ages younger than the current lower age anchor of $80$\,Myr set by \alper.

Although \alper is not in the LOPS2 field, both NGC 2451\,A and \alper are similarly aged with indistinguishable rotation sequences. As such we choose to model them as one cluster, in line with the approach taken by \citet{2023ApJ...947L...3B} when fitting gyrochronal models at a given age. At the age of NGC 2451\,A \& \alper, stars of $\sim0.8M_{\sun}$ are arriving at the ZAMS; however, at this mass there is still a notable spread of fast rotators (see NGC 2451\,A in Figure \ref{fig period dist}), making the slow sequence hard to identify. We therefore choose $0.85M_{\sun}$ as our lower limit for this cluster where the slow sequence is clearly identifiable. For NGC 2451\,B, the ZAMS gives a lower mass limit of $1M_{\sun}$. For the upper mass limit, although the upper bounds for a convective envelope is $\sim1.3M_{\sun}$, many of the clusters in this work have few-to-no detected rotation periods above $\sim1.2M_{\sun}$, thus for continuity we set the upper mass limit for all three clusters to $1.2M_{\sun}$. To ensure only single star evolution is accounted for, binary and higher order stars (see Section \ref{ssec binaries} for an outline of the identification process) are removed. For \alper, the sample was selected following the same binary filtration conducted and explained by \citet{2023AJ....166...14B}.\par

To accurately model the slow sequence, remaining single star fast rotators need to be removed. We utilize the \textit{Random sample consensus (RANSAC)} module from \texttt{SCIKIT-LEARN} \citep{scikit-learn}, running a maximum of 1000 trials using 50\% of the data per trial to fit a negative parabola (restricted second degree polynomial) to the data. A parabolic model was chosen as it can approximate the slow sequence, given the scatter in the data, whilst avoiding overfitting. Because stars reach the ZAMS on mass-dependent timescales, the slow sequence should be intrinsically smooth, with a characteristic rotation period at a given mass that is set by the extent to which a typical star has undergone angular momentum loss by that age. Previous studies that adopt models with higher degrees of freedom introduce localized features that lack clear physical interpretation and are likely driven by data sparsity rather than underlying stellar physics. The choice to enforce a negative parabola is also motivated by the physical structure of the slow sequence. We iteratively fit a polynomial to the data with identified outliers removed and the process repeated on the clean sample until convergence on the slow sequence, within 10 iterations for each cluster. After outliers had been effectively removed, an MCMC fit (5000 steps, 32 walkers \& 1000 step burn in) was run on the cleaned slow sequence using \texttt{EMCEE} \citep{2013PASP..125..306F} to return the best fit model and credible interval ranges for each cluster sequence. \par

\begin{figure}
     \centering
     \includegraphics[width=\linewidth]
     {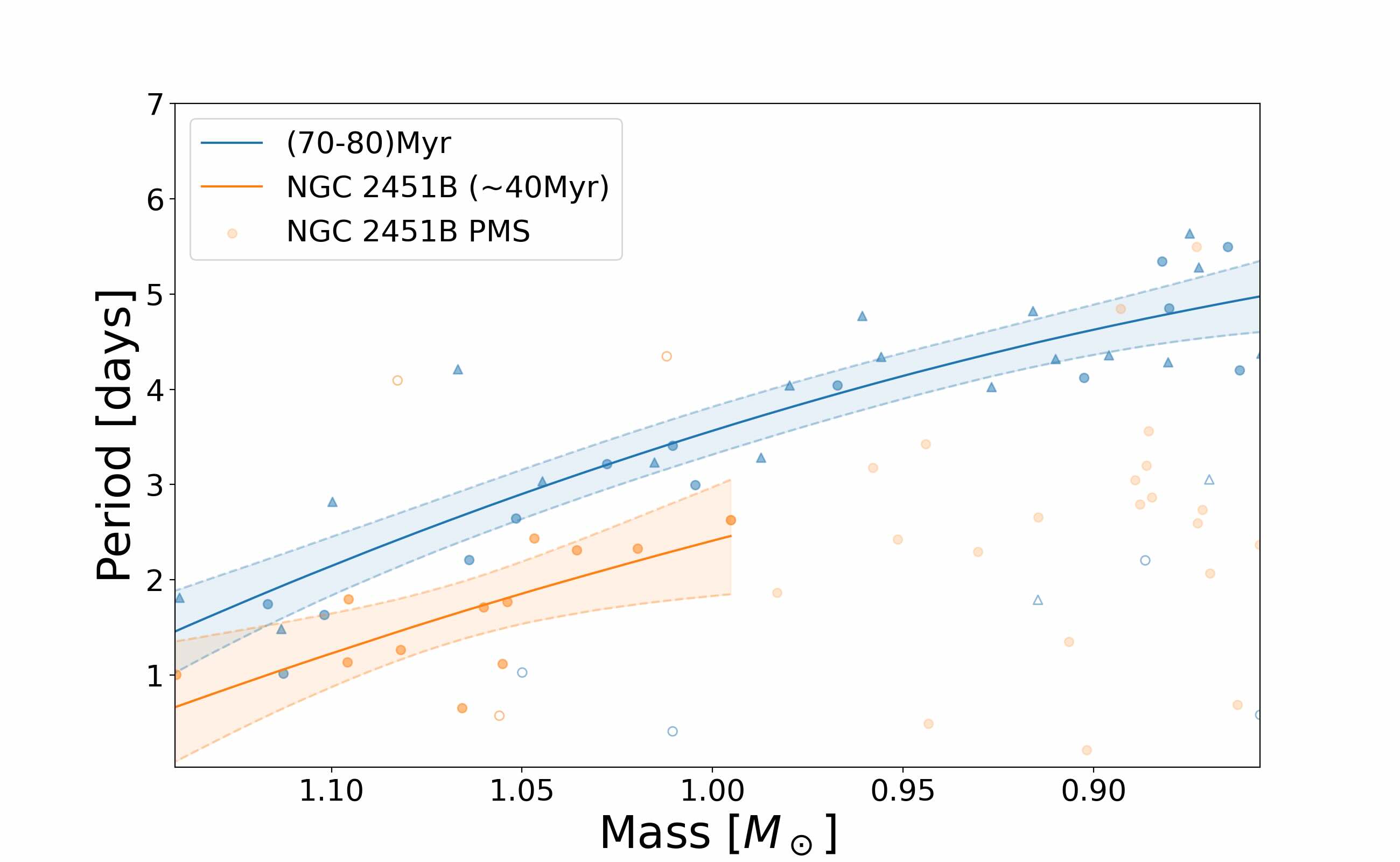}
     \caption{Best fit models and corresponding 95\% confidence intervals for the current lower gyrochronal age anchor of $\sim80$\,Myr and the slow sequence at $40$\,Myr. Blue triangles represent periods from \alper \citep{2023AJ....166...14B}, whilst blue circles represent periods from NGC 2451\,A. Open points represent outliers identified by the \textit{RANSAC} fitting process. NGC 2451\,B is shown in orange with PMS stars included with reduced opacity for completeness, but omitted from the modeling process.}
     \label{fig Li age MCMC}
 \end{figure}

The resulting slow sequence model and the corresponding 95\% confidence intervals are shown in Figure \ref{fig Li age MCMC}. NGC 2451\,A \& \alper exhibit a single well-defined slow sequence across the mass range $0.85 - 1.2M_{\sun}$, with periods increasing monotonically with stellar mass. Shaded regions represent the 95\% confidence intervals from the MCMC sampling, which shows good convergence of the sequence. Critically, both the best fitting slow sequence and its corresponding confidence interval for NGC 2451\,B is distinct from the canonically accepted lower age anchor of \alper used in existing empirically driven gyrochronology models. The distinct, narrow confidence intervals of NGC 2451\,B show the sequence is sufficiently well constrained to serve as a gyrochronal anchor for clusters aged at approximately solar mass ZAMS ages. Orange circles with reduced opacity ($<1\,\mathrm{M}_\odot$) in figure \ref{fig Li age MCMC} represent NGC 2451\,B members below the ZAMS mass limit for the clusters age, that are therefore still on the PMS in their spin up regime. Note that the increased spread of PMS stars in NGC 2451\,B is in stark contrast to the well constrained sequence of its main-sequence stars. This highlights the rate of convergence for fast rotating higher mass convective stars upon arrival at the ZAMS, making it possible to define the slow sequence at such a young age in this mass range. Moreover, whilst the stars in the PMS have not yet converged, the upper envelope of the distribution still sits below that of the $70-80$\,Myr slow sequence, which can further assist in relatively aging stellar groups below the current canonical $\sim$\,80 Myr gyrochronal age anchor. Utilising both the spread of rotation periods for a given mass bin and age in combination with the slow sequence location has also been employed recently by \citet{chronoflow}. \par

\subsection{Stalled spin down as a function of mass}
\label{ssec stall fn mass}

Stalled spin down, a phase of slowed or stalled spin down of the stellar surface, has previously been observed in young stars with prior studies typically focusing on the convergence of rotation periods at lower masses between $\sim0.7-1.4$\,Gyr \citep{2018ApJ...862...33A, sunteff, core_env_dynamo1}. During a period of stalled spin down, stars depart from the constant spin down rate proposed by \citet{1972ApJ...171..565S}, leading to faster than expected rotation periods at a given age and mass. This effect must be accounted for in order to accurately model the rotational evolution of young stars, however the cause of this slowed, or completely stalled spin down is still unclear. One possible explanation is the redistribution of angular momentum from the core to offset loss in the envelope through magnetised stellar winds \citep{ngc6811,2020A&A...636A..76S}. Accounting for this redistribution of angular momentum, \citet{2020A&A...636A..76S} recently modeled the evolution of the slow sequence, reproducing the period of stalled spin down between $\sim0.7-1.4$\,Gyr whilst also predicting that slowed or stalled spin down should be observed at different ages and mass ranges.\par

The net angular momentum exchange rate of the envelope can be summarised as in \citet{1991ApJ...376..204M};
\begin{equation}
    \frac{dJ_{\mathrm{env}}}{dt} = \frac{\Delta\mathrm{J}}{\tau_{\mathrm{c}}} - \frac{\mathrm{J}_{\mathrm{env}}}{\tau_{\mathrm{J}}}
    \label{eq dj_env}
\end{equation}

where $\mathrm{J}_{\mathrm{env}}/\tau_{\mathrm{J}}$ is the wind torque which follows an $\Omega^3$ dependence in the unsaturated regime \citep{1988ApJ...333..236K}. $\mathrm{J}_{\mathrm{env}}$ is the total angular momentum content in the convective envelope and $\tau_{\mathrm{J}}$ is the instantaneous exponential decay timescale for angular momentum loss via magnetic wind braking. \par
$\Delta\mathrm{J}/\tau_{\mathrm{c}}$ represents the coupling torque, with $\Delta \mathrm{J} \equiv \frac{\mathrm{I_{\mathrm{env}}}\mathrm{J_{\mathrm{core}}} - \mathrm{I_{\mathrm{core}}\mathrm{J_{\mathrm{env}}}}}{\mathrm{I_{\mathrm{core}}} + \mathrm{I_{\mathrm{env}}}}$ defined as the instantaneous amount of angular momentum required to equilibrate the rotational velocities of the core and envelope (i.e. satisfy solid body rotation; \citealt{1991ApJ...376..204M}). $\tau_\mathrm{c}$ is the core-envelope coupling timescale, the characteristic timescale to transfer this difference. $\Delta \mathrm{J}$ is a dynamic quantity that evolves through time. In contrast, $\tau_\mathrm{c}$, is held constant throughout the evolution of the star, representing a phenomenological simplification of how internal physical processes affect the rate with which angular momentum is redistributed within the stellar interior. A more physically motivated way to think of $\tau_\mathrm{c}$ is the intrinsic mass-dependent rate of angular momentum transfer between the core and envelope.

For stars $\gtrsim 1.2M_{\sun}$ formation of the core happens very quickly and encompasses almost the entire stellar interior, acting as a vast reservoir containing most of the total angular momentum content of the star. Stars in this mass range arrive at the ZAMS with high initial values of $\Delta\mathrm{J}$ and short $\tau_\mathrm{c}$ timescales; thus are readily able to replenish lost angular momentum in their thin convective envelopes on timescales comparable to the wind braking timescales. Stalled spin down therefore occurs early and because the core reservoir is so large, can be sustained over long timescales. This is observed in Figure \ref{fig period dist} whereby the rotation periods in mid-late F type stars show minimal rotational evolution across an approximately gigayear timescale. \par

Figure \ref{fig period dist} also shows that at solar masses, the slow sequence evolves from that of NGC 2451\,B ($\sim40$\,Myr) when these stars arrive at the ZAMS, to that of NGC 2451\,A ($\sim70$\,Myr), whereby there is then complete stalling of the sequence to at least the age of NGC 2516 ($\sim 150\,\mathrm{Myr}$). This shows that as mass decreases, a relatively smaller core leads to lower initial values of $\Delta\mathrm{J}$ at the ZAMS, along with a longer coupling timescale and consequently a lower overall initial coupling torque. Wind torques therefore dominate during early main-sequence evolution, explaining why low mass stars are seen to spin down relatively fast once they arrive on the ZAMS. Thus, due to the $\Omega$ dependence of the stellar wind stalling occurs later, at longer periods, when wind torques have been considerably weakened and a smaller coupling torque is required to offset wind braking (see Eq\ref{eq dj_env}). This has been shown in \cite{core_env_dynamo1}, who identified increased levels of activity in stalled K stars in Praesepe, which was attributed to a state of decoupling between the core and envelope.\par

Stalled spin down is therefore strongly mass dependent, whereby it occurs early on for higher mass stars with a subsequent wave of slowed or stalled spin down that propagates as both a function of mass and age. In other words, as mass decreases stars reach a level of $dJ_{\mathrm{env}}/\mathrm{dt} = 0$ at increasing ages. Smaller core reservoirs and comparatively large envelopes lead to lower initial $\Delta \mathrm{J}$ values and longer $\tau_\mathrm{c}$ that require extended time intervals for the wind torque to drive up $\Delta \mathrm{J}$ whilst simultaneously weakening according to its $\Omega$-dependence. This stalling wave can be inferred from the results of \cite{2023ApJ...947L...3B} who stated uncertainties in sun-like stars typically improve monotonically with age and the age posterior in K-dwarfs are highly asymmetric due to stalled spin down.

\subsection{The effect of stalled spin down on gyrochronology}
\label{ssec dga}

 \begin{figure*}
    \centering
    \includegraphics[width=0.7\linewidth]{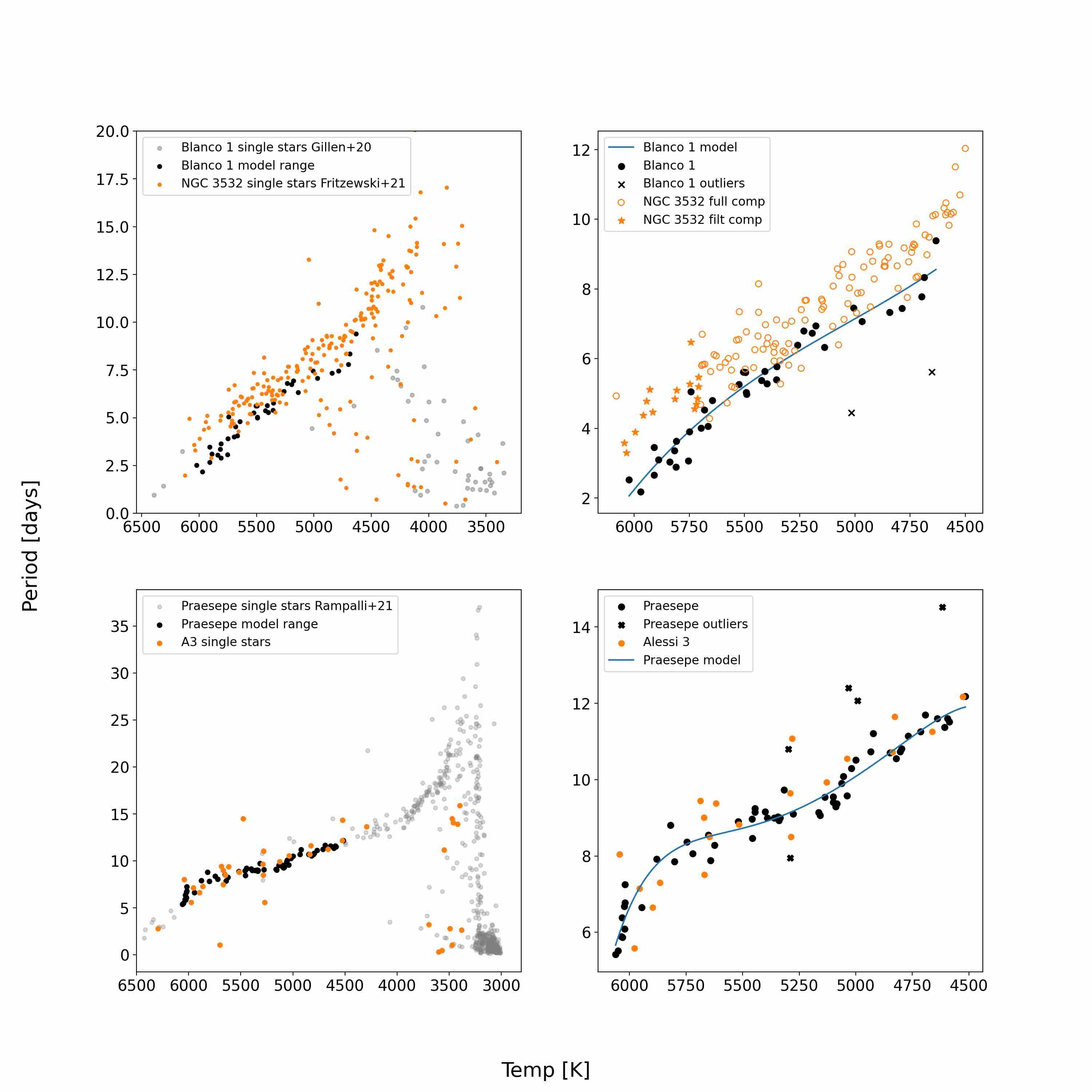}
    \caption{\textbf{Top row:} DGA comparison between NGC 3532 \citep{2021A&A...652A..60F} and Blanco 1 \citep{2020MNRAS.492.1008G}. Grey points represent the full single star distribution for Blanco 1, black points show the selected model range. The blue plot shows the modeled slow sequence for the model cluster (Blanco 1). Orange circles show the full single star sequence for NGC 3532. Open orange circles in the top right plot represent the full comparison range, with filled stars showing the subsequent filtered comparison range (see text). \textbf{Bottom row:} Same as above for Alessi 3 and Praesepe \citep{2021ApJ...921..167R}.}
    \label{fig a3 praesepe}
\end{figure*}

The mass and age dependence of stalled spin down effects the ability to which gyrochronal processes are useful in age dating unknown stars. As shown in section \ref{ssec lowering gyro limit}, the slow sequence at $40$\,Myr sits distinctly below that of the stalled sequence at $\sim70$\,Myr (see Figure \ref{fig Li age MCMC}). In this case, stalled spin down actually aids in age dating. During this window, an increasing fraction of stars migrate and settle onto the slow sequence, dragging the median fit progressively upward. The position of the sequence between these two boundaries therefore encodes relative age, enabling clusters $<70$\,Myr to be ordered chronologically by where their slow sequence fit falls.\par

Conversely, in older clusters the slow sequence has already converged and follows approximately the \citet{1972ApJ...171..565S} spin down relation. Any mass range now undergoing slowed or stalled spin down departs from this constant spin down relation, leading to younger than expected age estimates over the stalled mass range. This mechanism therefore has important consequences for gyrochronal processes such as differential gyrochronology aging (DGA) which has previously been used in the literature to infer cluster ages \citep{douglas19_praesepe,ngc6811,2023AJ....166...14B}.\par

DGA, the process of utilising a reference model cluster of well-constrained age to approximate the age of a second cluster, requires modeling of the reference cluster's slow sequence. The PMS lifetime is inversely proportional to mass with stars also requiring some time interval upon arrival at the ZAMS to converge onto the sequence. At younger ages, DGA is therefore restricted to roughly solar mass stars which stall very early. Moreover, lower mass stars both take longer to converge onto the slow sequence and experience stalling at later ages, meaning that DGA is most applicable precisely in the mass and age regimes where slowed, or stalled spin down is most likely to occur. Careful selection of the reference cluster and chosen mass comparison range (or some proxy for mass such as temperature) is therefore paramount.\par

To illustrate this point, we follow the DGA process outlined in section 6.3 of \cite{douglas19_praesepe} and equation 2 from \cite{2023AJ....166...14B} to calculate a differential age for NGC 3532, a benchmark cluster used to calibrate existing models, with an accepted age of 300\,Myr \citep{ngc3532_age1,ngc3532_age2,ngc3532_age3}. Using the rotation period distribution from \cite{2020MNRAS.492.1008G} for the 120\,Myr Blanco 1 as a reference cluster, gives two ages in which the mass ranges that have converged onto the slow sequence $(\gtrsim 0.7M_{\sun})$, thus are applicable for DGA, also largely coincide with a period of apparent stalled/slowed spin down $\sim (0.7-0.9M_{\sun})$.

We use the Sun to calculate the braking index, adopting a solar age of $4567$\,Myr \citep{sunage}, a solar period of $26.09\,\mathrm{d}$ \citep{sunper} and a solar temperature of $5789K$ (using the colour-effective temperature relation from \cite{sunteff} on the colour stated in \citet{douglas19_praesepe}). We model Blanco 1 between 4500K-6100K to ensure only stars on the main-sequence are considered whilst also avoiding the highest temperatures where there is a dearth of available rotation periods. For NGC 3532, we utilised the data from \cite{2021A&A...652A..60F}, including their binary filtration to filter for single stars.  Clear visual outliers that remained were manually cut before
fitting to ensure the polynomial closely tracked the converged slow sequence. The resulting comparative temperature range for NGC 3532 is represented by open orange circles in the top right panel of Figure \ref{fig a3 praesepe}. An age for NGC 3532 was calculated to be $156\,\mathrm{Myr} \pm 57\,\mathrm{Myr}$ (median age $\pm 1\,\sigma$), roughly half the canonical age used to calibrate gyrochronology models. If instead, we restrict our comparison range to $5700\,\mathrm{K}<\mathrm{T}_{\mathrm{eff}}<6050$\,K (temperatures that have between the ages of our reference and comparison clusters, already undergone stalling), this restricted sample of 16 stars (represented by filled orange stars in Figure \ref{fig a3 praesepe}) returns an age for NGC 3532 of $255\pm83$\,Myr, accurate to within errors of the currently accepted literature age for the cluster.\par 

If we now turn our attention to Alessi 3 where, over the original temperature range used above stalling has ceased, we can show how the same range (4500K-6100K) provides a good comparison for differential aging. From our literature review, current age estimates for Alessi 3 range from approximately $(400 - 800)$\,Myr. Using the gyrochrones from \citet{2023ApJ...947L...3B}, we can see that the rotation periods in this temperature range show clear evolution across the entire age estimate range of Alessi 3. The rotation period profiles of Alessi 3 and Praesepe are, however, almost indistinguishable. Thus we conclude that stars in the comparison range should not be stalling but instead the two clusters are approximately coeval.
Using the results from \citet{2021ApJ...921..167R}, we compare Alessi 3 to Praesepe in the same manner as our analysis of NGC 3532. Praesepe data was filtered for binary stars using the binary flag in \citeauthor{2021ApJ...921..167R}, along with any target with a RUWE value $> 1.4$. This process gave a sample of 57 stars to create our model. Using a sample of 20 stars inside the comparison range, we derive a differential age for Alessi 3 of $687\,\mathrm{Myr} \pm 106\,\mathrm{Myr}$, accurate to the age range given by our literature review and approximately coeval with that of Praesepe.\par

Thus, whilst DGA can be a powerful tool for estimating cluster ages, these results highlight the importance of the mass range (or some proxy) considered in the comparison. At a given age, careful selection of the reference cluster utilized is essential to ensure the two clusters have adequate sample stars not undergoing slowed or stalled spin down to provide a meaningful comparison. \par

\section{Conclusions}
\label{sec conc}

We conducted a photometric study of seven young $(< 1\,\mathrm{Gyr})$ open clusters that lie within the upcoming PLATO LOPS2 field; namely, Alessi 3, NGC 2516, NGC 2451\,A \& B, Trumpler 10, IC 2391 and Collinder 135. We report 1063 rotation periods in total, of which 479 are newly analysed in this cluster specific rotation study and 63 are unique periods not previously reported in the literature. The distributions for Alessi 3, Trumpler 10 \& NGC 2451\,B represent the first known dedicated photometric studies on rotation for these clusters, providing valuable insights at important stellar evolutionary stages. We also conduct our own identification of higher order systems, identifying both photometric and astrometric binaries over the RUWE threshold of 1.4, identifying 285 higher order systems from our measured rotation period distributions across all seven clusters. We find good agreement with the results of this work and those of previous photometric studies on the observed clusters.\par

\cite{2023ApJ...947L...3B} recently improved current gyrochronology models, anchoring the lowest age to that of \alper at 80\,Myr. We compare the youngest cluster in this work with a well constrained age estimate, NGC 2451\,B $(43\pm3\,\mathrm{Myr})$, to the canonically accepted lower age anchor for gyrochronology to see if the slow sequences undergo discernible evolution over this age difference. Following etiquette by previous studies, we treat \alper and NGC 2451\,A as one large cluster due to their similar ages and indistinguishable rotation profiles, comparing it to the evolutionary state of the younger NGC 2451\,B by modeling their respective slow sequences. The resulting best fit sequences show a clear distinction between that of NGC 2451\,B at $\sim40$\,Myr and the sequence at $\sim70-80$\,Myr, showing that it is possible to identify cluster ages at roughly half the currently accepted lower age limit. For these F-type stars, we observe the rapid convergence of points onto a well defined slow sequence and subsequent stalling between the ages of $\sim70$\,Myr to at least 150\,Myr. For clusters aged between $40-70$\,Myr the level of convergence of stars onto this stalled sequence, given the amount of time the stars have had to spin down, therefore encodes the relative age of the cluster. We note however, the low number of available rotation periods for solar-mass stars. Thus, whilst this process shows real promise in lowering the gyrochronal age anchor, it highlights the importance of the PLATO mission \citep{2024arXiv240605447R}, which will observe the LOPS2 field, encompassing NGC 2451 B and other young clusters, for a minimum of two years. These photometric observations will be vital in helping to drive forward our understanding of young stellar evolution by increasing the sample size of measured rotation periods across a range of both mass and age.\par

At lower masses we see a period of spin down before slowed, or even completely stalled evolution, in line with increased $\tau_\mathrm{c}$ values. This is in good agreement with the predictions of \cite{2020A&A...636A..76S} and is evidence for the dynamic relationship between core-envelope coupling and magnetic braking being a dominant factor in the process of stalled spin down. The recent results by \cite{core_env_dynamo1} highlights increased levels of surface activity during stalling at the age of Praesepe resulting from core-envelope decoupling. Due to the winds $\Omega-$dependence, at longer periods the wind torque weakens dramatically and thus lower overall levels of core envelope coupling are required to offset angular momentum loss rates by the wind, further supporting the idea of stalling resulting from angular momentum redistribution from the core offsetting losses by magnetised stellar winds.\par

Finally, following the process previously used to age date open clusters, we utilise differential gyrochronology aging (DGA) to age date Alessi 3, providing a new age estimate of $687\pm106$\,Myr. We also show the importance of selecting similarly aged clusters with an adequate mass range not undergoing slowed or stalled spin down, where comparing stalled regions of the period-mass space have a detrimental effect on age estimates and their uncertainties attained via DGA.  

\section*{Acknowledgements}
AH thanks Victor See and Guy Davies for their enjoyable and informative discussions. We would also like to thank the anonymous referee for their careful reading of the manuscript and insightful suggestions for improvement. AH gratefully acknowledges support from UK Research and Innovation (UKRI) via a training grant awarded by the Science and Technology Facilities Council (STFC) (grant number ST/X50869X/1). EG, MPB and DC gratefully acknowledge support from UK Research and Innovation (UKRI) under the UK government’s Horizon Europe funding guarantee for an ERC starting grant (grant number EP/Z000890/1).
JSJ gratefully acknowledges support by FONDECYT grant 1240738 and from the ANID BASAL project FB210003. This research is based on data collected under the NGTS project at the ESO La Silla Paranal Observatory. The NGTS facility is funded by a consortium of institutes consisting of the University of Warwick, the University of Leicester, Queen’s University Belfast, the University of Geneva, the Deutsches Zentrum f\"ur Luftund Raumfahrt e.V. (DLR; under the ‘Großinvestition GI-NGTS’), together with the UK Science and Technology Facilities Council (STFC; project references ST/M001962/1, ST/S002642/1 and ST/W003163/1). This research has made use of the VizieR catalogue access tool, CDS, Strasbourg, France (DOI : 10.26093/cds/vizier). The original description of the VizieR service was published in A\&AS 143, 23 (\cite{2000A&AS..143...23O}). This research has made use of the SIMBAD database, operated at CDS, Strasbourg, France(\cite{2000A&AS..143....9W}). This work has made use of data from the European Space Agency (ESA) mission Gaia (\url{https://www.cosmos.esa.int/gaia}), processed by the Gaia Data Processing and Analysis Consortium (DPAC, \url{https://www.cosmos.esa.int/web/gaia/dpac/consortium}). This paper includes data collected by the TESS mission that are publicly
available from the Mikulski Archive for Space Telescopes (MAST).

\section*{Data Availability}

Supplementary data are available at MNRAS online.\par
\noindent \textbf{Table 1.} Results and supplementary metadata for all stars with rotation periods.


\bibliographystyle{mnras}
\bibliography{bibli} 



\appendix

\section{Cluster membership table: column overview}
\label{app cl memb tab}
\newpage
\begin{table*}
\centering
\begin{tabular}{|c|l|}
\hline
\hline
\textbf{Column header} &
  \textbf{Description} \\ \hline
Cluster &
  Canonically accepted cluster name. \\ \hline
\begin{tabular}[c]{@{}c@{}}Cluster\\ centre (l)\end{tabular} &
  Galactic longitude of cluster centre (Degrees) as cited by \citet{Cantat-Gaudin2020}. \\ \hline
\begin{tabular}[c]{@{}c@{}}Cluster\\ centre (b)\end{tabular} &
  Galactic latitude of cluster centre (Degrees) as cited by \citet{Cantat-Gaudin2020}. \\ \hline
Distance &
  Distance to cluster (pc) as cited by \citet{Cantat-Gaudin2020}. \\ \hline
\# Cands. full &
  filtered candidacy list, whereby a star is referenced as a member of the cluster in at least one of the accepted literature review studies. \\ \hline
\# Membs. full &
  Number of members after enforcing minimum probability of membership $\geq70\%$ on \# Cands. full. \\ \hline
\# Membs. filt &
  \begin{tabular}[c]{@{}l@{}} Number of members after removal of \citep{Kounkel-Covey2019, Meingast2021}. \\ (This membership number is typically more representative of the canonically accepted cluster name of column 1).\end{tabular} \\ \hline
\# Membs. with NGTS LC’s &
  Number of members for which an NGTS lightcurve was extracted. \\ \hline
Obs dates &
  Dates for which NGTS observations on the cluster were undertaken (yyyy/mm/dd) \\ \hline
\end{tabular}
\caption{Overview of column headers for table \ref{tab: cl overview}.}
\label{app tab table 1 overview}
\end{table*}

\newpage
~
\newpage

\section{Literature Period comparison overview}
\label{app pcomp}
\subsection*{NGC 2451\,A}
Of 62 stars in common with \citet{2024ApJ...962...16D}, three targets (TIC 174661301, TIC 174423244 \& TIC 175173310) showed disagreement, all consistent with half/double period aliases. For the former two, the rotation periods in this work are roughly twice that reported by \citeauthor{2024ApJ...962...16D} (6.98\,d \& 7.22\,d compared to 3.47\,d \& 3.54\,d, respectively). In both cases, the period identified by NGTS data is a clear significant peak in the periodogram whereas that reported by the literature is not observed at all. Cross-examination of TESS light curves used by \citeauthor{2024ApJ...962...16D} (CDIPS \& QLP, sectors 6-11) showed that either the maximum periodogram peak actually agreed with the period in this work, or exhibits evidence of a double dip structure. We conclude that the true period is that identified by NGTS here. For TIC 175173310, we identify half the period of \citeauthor{2024ApJ...962...16D} (0.167d instead of 0.33\,d). Although TESS shows a clear modulation on a 0.33\,d signal, NGTS data show a better fit to 0.167\,d. We note however that 0.33\,d is a strong alias period for the diurnal nature from ground-based observations and thus, if 0.33\,d is the true period, systematics may have affected our ability to detect this and we may be reporting half the true period. \par

\subsection*{Collinder 135}
Of the 35 stars in common with \citeauthor{2024ApJ...962...16D} in Collinder 135 there are only two discrepancies, both consistent with half-period harmonics. 
We measure a period of 0.45d for TIC 22640107 whereas \citeauthor{2024ApJ...962...16D} report 0.29d. CDIPS data of Sector 7 in TESS, used in \citet{2024ApJ...962...16D}, shows increased activity and two strong periods (the highest Lomb-Scargle periodogram peak lies at the longer period found in this work and a second, slightly lower peak, at the shorter period cited by \citeauthor{2024ApJ...962...16D}. Additional TESS sectors show clear modulation on the longer 0.45d period, which we conclude is the true period. The same is true of TIC 23093276, in which only CDIPS data from Sector 7 appears to have been available, but probing later TESS sectors again confirms the period assigned using NGTS data. \par

\subsection*{IC 2391}
IC 2391 presents both the most outliers and also the largest relative difference between periods of the three clusters. TIC 93833881 \& TIC 93549861 can be explained much like most other discrepancies, with NGTS showing a clear period on a 2:1 harmonic with TESS data. Reviewing TESS data, we conclude the true periods are that reported by NGTS here. The largest disparity in period is from TIC 145667924. The literature period is 9.54d whereas the returned period by this work is 0.19d, a $\sim 50\times$ difference. Both periods are observed by NGTS and TESS, however in the available TESS sectors (8-9) for QLP \& CDIPS the primary peak is the longer period. Conversely in NGTS data the shorter 0.19d period is the highest power peak, with the longer period being substantial but lower power. We cite the highest power period in this work, but advise caution when interpreting this target. \par

The final outlier, TIC 93630218 has a literature period of 0.2d. This period is not observed at all in NGTS data. Observing TESS light curves a signal around the NGTS period of 3.6d is observable within the shorter period oscillations, however the relative power of this peak is heavily diluted by the large amplitude modulation on the shorter period. TIC 93630218 is dominated by contaminant flux, with a TESS contamination ratio of $1.7$. With the NGTS contamination ratio below our threshold value, it is likely the true period is that of the NGTS signal and the TESS assigned period is from a brighter contaminating source.

\subsection*{NGC 2516}
\label{sssec ngc2516}

\citet{2021AJ....162..197B} provides the most rotation periods of the clusters considered in this work, investigating 3298 candidate members within the core and halo, with 987 light curves meeting the initial cleaning criteria as outlined in their paper (referenced as \textit{setA} in their study). \citet{2020A&A...641A..51F} conducted their own membership calculation with 844 stars meeting their membership threshold for the core of NGC 2516, returning 308 rotation periods from their membership list. In order to determine a hierarchy for which analysis should take precedence for stars with multiple cited periods, we compared the results of shared periods between each study with this work, calculating the fractional period difference for each shared period. We reason that, due to the long baseline observations coupled with high levels of photometric precision, NGTS is well suited to photometric rotation studies. Therefore, the study with periods closer aligned to those in this work will more accurately represent the full rotation distribution of NGC 2516. Whilst good agreement is generally found with both studies, we find better agreement with the results of \citeauthor{2021AJ....162..197B}, with a median fractional difference of $0.055$ compared to $0.136$ with \citeauthor{2020A&A...641A..51F} and therefore assign it a higher precedence for shared periods in the literature. \citet{2007MNRAS.377..741I} is not considered in this comparison, as the overlap of rotation periods is restricted due to their deeper observations primarily observing low mass stars. Figure \ref{fig period comp} shows that the periods attained in this work agree largely with those cited in the literature to date. In both cases, there are few outliers from the 1:1 and period harmonic tracks. For such targets, a period comparison was conducted by phase folding the NGTS photometry on both the period assigned during this analysis and the literature period with the folded data and detrended periodogram then being reviewed by eye. \par

In two cases (TIC 358464823 \& 308307537) it was decided that the period reported by the literature was correct and the period cited initially by this work was half the true rotation period. For both targets, the period was changed to reflect this \footnote{N.b. We still report a period attained by analysis of NGTS photometry. In both cases the second highest power peak in the periodogram closely reflects the literature period and thus, we report the corresponding period.}. \par

\begin{figure*}
    \centering
    \includegraphics[width=\linewidth]{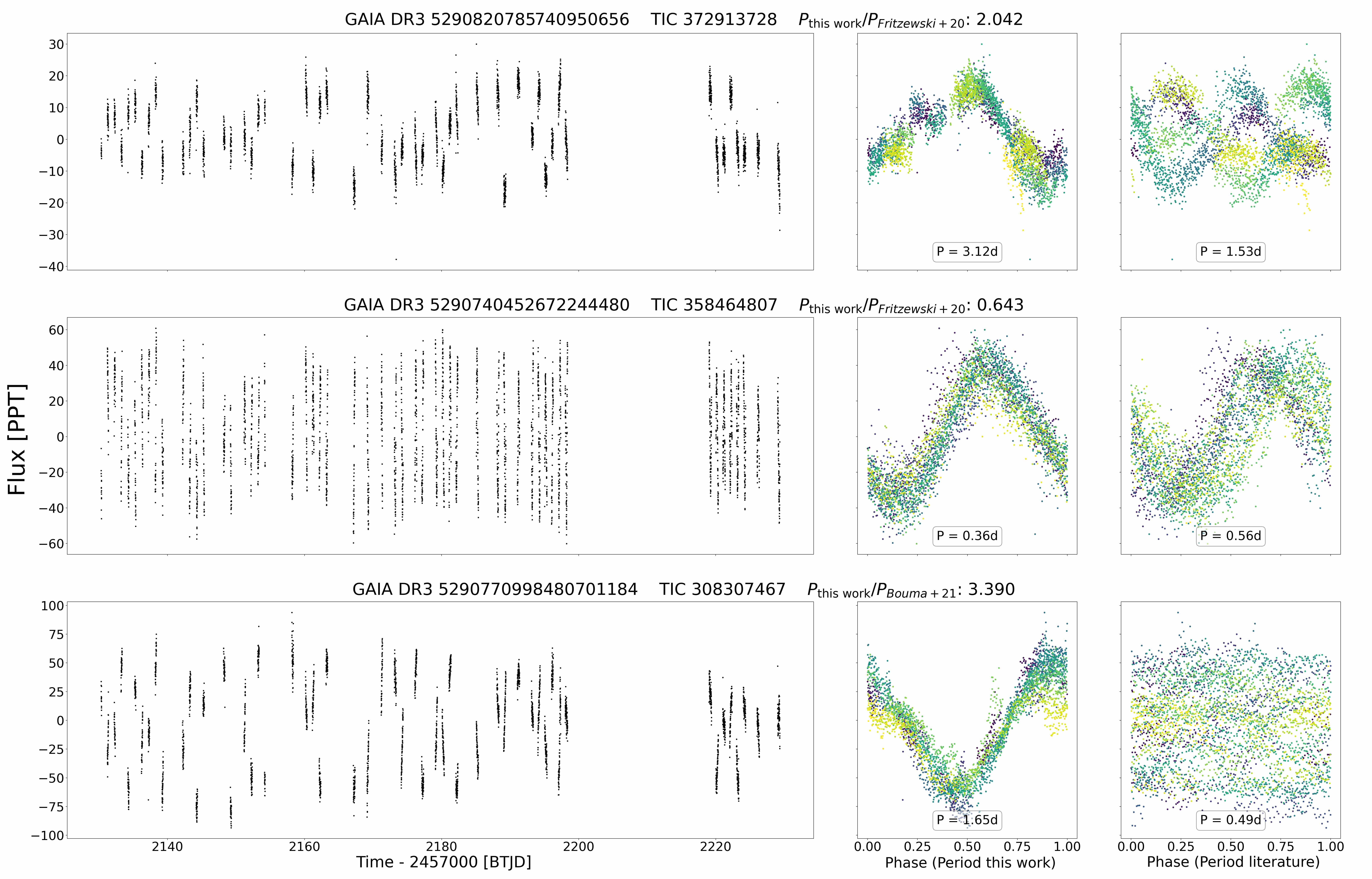}
    \caption{Examples of NGTS light curves, phase folded on both the adopted period from this work and the period cited by either \citet{2020A&A...641A..51F} or \citet{2021AJ....162..197B}.}
    \label{app diagnostic}
\end{figure*}

In all other cases we conclude the NGTS cited period here is the true period. In many cases, we did not observe a clear periodic signal on the literature period at all. Sometimes, a peak was observed at the literature cited period, but these were typically harmonics of the main period identified by this work (see Figure \ref{app diagnostic} for examples of NGC 2516 light curves and the corresponding phase fold for the period cited by this work and that from the literature).

\subsection*{TESS All-Sky Rotation Survey (TARS)}

The final panel of Figure \ref{fig period comp} shows the period comparison between this work and \citet{2026arXiv260305586B}, showing good agreement across all seven clusters. Interestingly, \citeauthor{2026arXiv260305586B} updated their catalog after reviewing their period-doubling logic to account for stars exhibiting double dip features. Version one (prior to this update) showed a larger discrepency with a substantial number of targets in TARS reporting double the period to this work. Version two (post update) finds much better agreement to this work. Of the 1000 shared periods, 894 targets have adopted rotation period values to within a 10\% difference \& 686 targets within 2\%. This correction by \citeauthor{2026arXiv260305586B} and its subsequent improvement in agreement between our results further validates the ability of long baseline NGTS photometry to discern the correct rotation period from the complex light curves of young stars.\par

Following the process outlined in Section \ref{sssec ngc2516}, we compare the adopted rotation periods between the two studies by phase folding the period on the detrended NGTS photometry for all stars with rotation period discrepancies $>10\%$. For three stars (TIC 174797693, 382579084 \& 364398130), we determined the rotation period cited by \citeauthor{2026arXiv260305586B} to be the correct period and that adopted initially in this work was an alias period incorrectly identified as the true period. For all other discrepancies, we conclude that the rotation period adopted by this work is the correct period (see appendix \ref{app tars} for examples).

\begin{figure*}
    \centering
    \includegraphics[width=\linewidth]{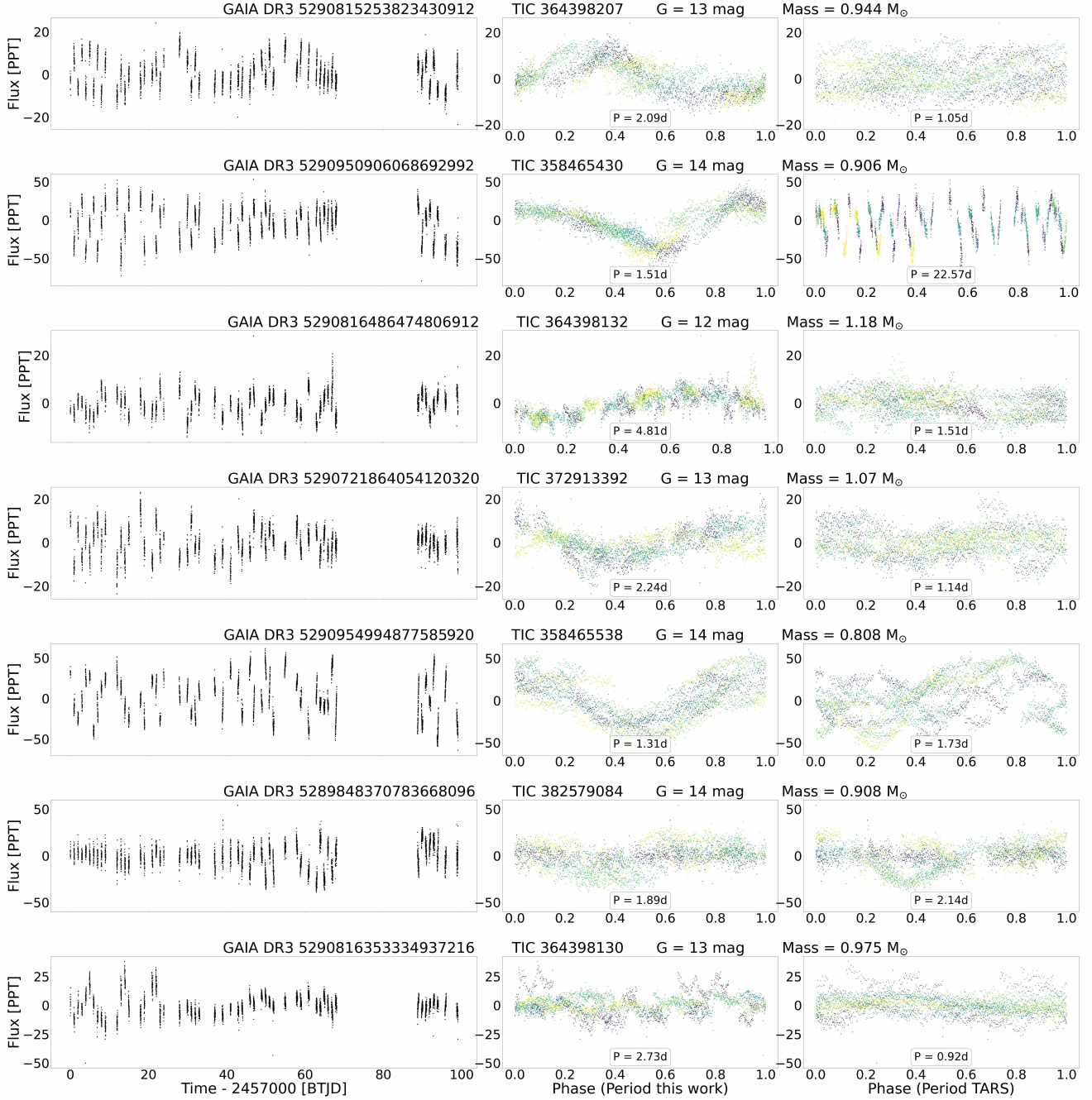}
    \caption{Examples of period discrepancies between this work and \citet{2026arXiv260305586B}. For the top five systems we adopt our period while for the bottom two panels, we determine that the period reported by \citeauthor{2026arXiv260305586B} is the correct period and reflect this in our results.}
    \label{app tars}
\end{figure*}

\newpage
~
\newpage
~
\newpage
~
\newpage
\section{Data summary table format}
\begin{table}
\centering
\caption{Column headers and descriptions for data summary table. This table is available in its entirety in machine readable format from the online journal.}
\begin{tabular}{|c|l|}
\hline\hline
\textbf{Column header} &
  \textbf{Description} \\ \hline
GAIA\_DR3 &
  Gaia DR3 identification. \\ \hline
tic\_id &
  TESS TIC identification. \\ \hline
Period &
  Adopted period in days. \\ \hline
gmag\_phot\_bin &
  \begin{tabular}[c]{@{}l@{}}Photometric binaries identified via CMD fitting (see Section 3.3).\\ (Boolean identifier, True == bin)\end{tabular} \\ \hline
ruwe &
  Gaia DR3 ruwe values. \\ \hline
binary\_1.2 &
  \begin{tabular}[c]{@{}l@{}}If star is identified as either photometric or ruwe \textgreater{}1.2 \\ (Boolean identifier, True == bin)\end{tabular} \\ \hline
binary\_1.4 &
  \begin{tabular}[c]{@{}l@{}}If star is identified as either photometric or ruwe \textgreater{}1.4 \\ (Boolean identifier, True == bin)\end{tabular} \\ \hline
Teff\_DR3 &
  Gaia DR3 cited Teff in Kelvin. \\ \hline
Teff\_DR2 &
  Gaia DR2 cited Teff in Kelvin. \\ \hline
G\_mag &
  \begin{tabular}[c]{@{}l@{}}Gaia DR3 Gmag (Absolute).\\ Detrended using process outlined in section 3.3\end{tabular} \\ \hline
bp\_rp\_0 &
  \begin{tabular}[c]{@{}l@{}}Dereddened colour index.\\ Detrended using process outlined in section 3.3\end{tabular} \\ \hline
gmag &
  Gaia DR3 Gmag (apparent) \\ \hline
bp &
  Gaia DR3 BP passband value. \\ \hline
rp &
  Gaia DR3 RP passband value. \\ \hline
Plx &
  Gaia DR3 parallax value. \\ \hline
e\_parlx &
  Gaia DR3 parallax error. \\ \hline
MIST\_mass\_v0 &
  \begin{tabular}[c]{@{}l@{}}Stellar mass calculated using MIST EEP tracks (inital v/vcrit = 0), in solar mass. \\ Calculated where possible from Gaia DR3 Teff, else DR2 Teff\end{tabular} \\ \hline
MIST\_mass\_v4 &
  \begin{tabular}[c]{@{}l@{}}Stellar mass calculated using MIST EEP tracks (initial v/vcrit = 0.4), in solar mass. \\ Calculated where possible from Gaia DR3 Teff, else DR2 Teff\end{tabular} \\ \hline
cl\_nm &
  Parent cluster name. \\ \hline
\end{tabular}
\label{tab: results table}
\end{table}


\bsp	
\label{lastpage}
\end{document}